\documentclass[11pt,a4paper]{article}
\usepackage{jcappub}
\usepackage[british]{babel}
\usepackage{latexsym,amsmath,amssymb}
\usepackage{graphicx}
\usepackage{booktabs}
\usepackage{hyperref}

\def\LL{{\cal L}}
\def\MM{{\cal M}}

\def\dd{\partial}
\def\be{\begin{equation}}
\def\ee{\end{equation}}
\def\bea{\begin{eqnarray}}
\def\eea{\end{eqnarray}}
\def\ba{\begin{array}}
\def\ea{\end{array}}
\newcommand{\lsim}{\,\raise 0.4ex\hbox{$<$}\kern -0.8em\lower 0.62ex\hbox{$\sim$}\,}
\newcommand{\gsim}{\,\raise 0.4ex\hbox{$>$}\kern -0.7em\lower 0.62ex\hbox{$\sim$}\,}

\def\de{\mathrm{DE}}
\newcommand{\ca}{{c_a^2}}
\newcommand{\cs}{{c_s^2}}

\newcommand{\brho}{\bar{\rho}}
\newcommand{\dep}{\delta p}
\newcommand{\der}{\delta\!\rho}

\newcommand{\HH}{{\mathcal H}}

\newcommand{\eff}{{\mathrm{eff}}}
\newcommand{\nad}{{\mathrm{nad}}}
\newcommand{\rel}{{\mathrm{rel}}}
\newcommand{\cseff}{{\hat{c}_{s,\mathrm{eff}}^2}}

\newcommand{\cc}{{\delta_{_0}}}

\newcommand{\mpl}{{M_\mathrm{Pl}^2}}

\author{Martin Kunz}
\affiliation{D\'epartement de Physique Th\'eorique and Center for Astroparticle Physics, Universit\'e de Gen\`eve, Quai E.\ Ansermet 24, CH-1211 Gen\`eve 4, Switzerland}

\emailAdd{martin.kunz@unige.ch}

\begin{document}

\title{The phenomenological approach to modeling the dark energy}

\abstract{
In this mini-review we discuss first why we should investigate cosmological models
beyond $\Lambda$CDM. We then
show how to describe dark energy or modified gravity models in a fluid language with the help 
of one background and two perturbation quantities. We review a range of dark energy models
and study how they fit into the phenomenological framework, including generalizations like
phantom crossing, sound speeds different from $c$ and non-zero anisotropic stress, and how
these effective quantities are linked to the underlying physical models. We also discuss the
limits of what can be measured with cosmological data, and some challenges for the framework.
}




\maketitle  

\section{Introduction}  \label{sec:intro}

2011 has been an exciting year for dark energy: In the same week in early October, the Nobel prize in physics was given
to Saul Perlmutter, Brian P. Schmidt and Adam G. Riess {\em ``for the discovery of the accelerating expansion of the Universe through observations of distant supernovae''} and the {\sc Euclid} satellite mission\footnote{Continuously updated information on {\sc Euclid} is available on \url{http://www.euclid-ec.org}.} \cite{EditorialTeam:2011mu}
was selected by
ESA for the second Cosmic Vision launch slot (expected for 2019 -- but some delays are likely
between now and then). Dark energy has arrived in the main stream of physics, not only
cosmology, and it is seen as one of big challenges facing the discipline at the beginning of the
21st century. And at least observationally it really is a 21st century problem. While theoretically the cosmological
constant has been around since the development of General Relativity \cite{Einstein:1915ca,Einstein:1917ce} --
as the unique additional term that can
self-consistently be added to the Einstein equations, and because it would have permitted the construction
of a static universe -- the two observational publications for which the Nobel prize was given date from 1998 and 1999
respectively \cite{Riess:1998cb,Perlmutter:1998np}. 
Even though there were observational hints before then, it was really the evidence
presented in these two papers that led, in a way reminiscent of a phase transition, to the
general acceptance amongst cosmologists that some kind of {\em Dark Energy} was necessary.

There are several excellent reviews discussing dark energy in great detail, for example
\cite{Carroll:2000fy,Copeland:2006wr,Durrer:2008in,Brax:2009ae,Jain:2010ka,Sapone:2010iz,Clifton:2011jh} just to name a few, and
there is also a new book by Luca Amendola and Shinji Tsujikawa on the topic \cite{2010deto.book.....A}.
In this mini-review, loosely based on an overview talk given at PONT 2011 in Avignon, 
I do not plan to compete with these weighty tomes and to discuss in depth
the dark energy in all its guises and with all its consequences. Instead, I will present a rather
personal view of the dark energy, focused on its phenomenological aspects, and I will be
trying to add a few ideas that in my opinion are interesting but a bit off the trodden path.
Many more aspects of the accelerating universe are discussed in the companion papers
to this one: {\em Observational Evidence of the Accelerated Expansion of the Universe} by
Pierre Astier and Reynald Pain \cite{cras:astier2012}, 
{\em Everything You always Wanted to Know about the Cosmological Constant (but Were Afraid to Ask)}
by J\'er\^ome Martin \cite{cras:martin2012}, {\em Establishing Homogeneity of the Universe in the Shadow of Dark Energy} by
Chris Clarkson \cite{cras:clarkson2012} and {\em Galileons in the Sky} by Claudia de Rham \cite{cras:derham2012}.

I will start in section \ref{sec:notcc} with the question whether the data is really reliable and needs to be taken seriously,
why it makes sense to go beyond a cosmological constant, and what other explanations
beyond $\Lambda$ we can think of.

In section \ref{sec:pheno} we will then introduce the phenomenological framework that describes the most general
fluid in terms of the variables present in the energy-momentum tensor. These variables allow
for general departures from $\Lambda$CDM and in addition they allow to understand in a precise,
quantitative sense, which degrees of freedom can be probed by cosmological measurements.

We review actual dark energy models in section \ref{sec:DE}, illustrating how they fit into the
framework and how different types can be distinguished through the values taken by their fluid parameters. We
also discuss how the phenomenological variables allow for straightforward generalizations
of the models (for example to different sound speeds) and how this is then reflected in ``model space''.

In the penultimate section \ref{sec:problems} we discuss some of the challenges and limits of the
phenomenological description, including the dark degeneracy which illustrates a fundamental
limitation of the framework that is a reflection of fundamental limitations of cosmological
measurements when trying to reconstruct the physics of the different constituents of the
Universe. We conclude this mini-review with a summary and an outlook on future developments in
section \ref{sec:conclusions}.

\section{Initial questions}  \label{sec:notcc}

\subsection{Reliability of the data}  \label{sec:data}

The first question one encounters when thinking about the dark energy is whether the observational
evidence is really compelling, or whether the best explanation is that the data is simply wrong. 
The current observational situation being discussed in detail elsewhere in
this volume \cite{cras:astier2012}, I will here just highlight a single point, the use of distance duality to check the internal 
consistency of the distance data. The distance data is important since it provides currently the most
precise measurements that only rely on the behavior of the average, homogeneous metric. It is also
of historic interest since it is this type of data that garnered the 2011 Nobel prize in physics.

In any metric theory of gravity, the distances are essentially unique, thanks to a theorem
proved in 1933 by Etherington that shows that for this class of theories the relation between luminosity distance
and angular diameter distance,
\be
\frac{d_L(z)}{d_A(z)} = (1+z)^2
\ee
(where $z$ is the redshift) always holds \cite{1933PMag...15..761E}. This means that we can test
the luminosity distance data with the help of angular diameter distance data, and vice versa
\cite{Bassett:2003vu,2004ApJ...607..661B,Uzan:2004my}. Since the two types of data agree
well even for today's precise measurements \cite{Avgoustidis:2010ju} this limits a wide range of systematic
effects that affect only one of the two distances. Examples on the luminosity distance side
include explanations for the observed dimming of supernovae like replenishing grey dust
or ultralight axions \cite{Kunz:2004ry}. More generally speaking, it is very difficult to
explain the supernova data by doing unpleasant things to the photons, since this would
generically break the distance duality. This leaves effectively only the option to change the
behavior of the metric, which is of course exactly what dark energy models do (but see also
\cite{Clarkson:2011br} for interesting conclusions on the metric with the help of
distance duality). For this reason it is very unlikely that the distance data is ``just'' wrong.

\subsection{Why go beyond $\Lambda$?}  \label{sec:lambda}

The cosmological constant dates back right to the development of General Relativity and has been
in and out of favor several times during the last century. But, as we will see later on, the data
is fully compatible with the dark energy being a cosmological constant, so why should we
bother to look elsewhere? There is a whole review in this volume
on the cosmological constant \cite{cras:martin2012}, so I will only mention here briefly the two main objections,
the naturalness problem and the coincidence problem. The first one appears because all particles should
give rise to quantum mechanical zero-point fluctuations $\sim \sqrt{k^2 + m^2}$ (with the fermions
having the opposite sign to the bosons). Since observations
indicate that the energy scale of the cosmological constant would have to be of the order of
$10^{-3}$ eV, we can integrate out the heavier particles like electrons. The integral of the
zero-point fluctuations over momentum $k$ results in
huge contributions to the total cosmological constant (that can in principle be absorbed into
a renormalization of a bare cosmological constant) as well as a logarithmic term (see 
e.g.~\cite{Maggiore:2011hw,Hollenstein:2011cz} for a recent discussion). Especially the
logarithmic contribution is highly problematic: Such terms should be considered physical as
they have been observed for masses and couplings, but their contribution to the vacuum energy, already from supposedly
well-known particles like the electron, are far larger than the observed value of the cosmological
constant. The one known mechanism to suppress the contribution from the zero-point fluctuations,
supersymmetry (which relates fermions and bosons to each other), is necessarily broken below a TeV or
else we would already have observed its effects with the help of particle accelerators.

The second issue, the coincidence problem, comes from the fact that if we look at the relative energy density in
the cosmological constant during the evolution of the Universe, $\Omega_\Lambda(a) = \rho_\Lambda/\rho_{\rm tot}$,
we find that there is a sharp transition around today where $\Omega_\Lambda(a)$ increases from
zero to one. It appears very unlikely that we are living just {\em during} this rapid transition, rather
than earlier or later. Let us illustrate the coincidence problem a bit more.
Writing just $\Omega_\Lambda$ for $\Omega_\Lambda(a=1)$ we find that close to today this 
function behaves as
\be
\Omega_\Lambda(a) \approx \frac{\Omega_\Lambda}{\Omega_\Lambda + (1-\Omega_\Lambda) e^{-3 N}} ,
\ee
where $N=\ln(a)$ is the number of e-foldings relative to today, and $\Omega_\Lambda\approx 0.73$.
Let us call the transition the period from $\Omega_\Lambda(a) = 0.05$ to
$\Omega_\Lambda(a) = 0.95$. We find that this corresponds to $N \in [-1.31,0.65]$, i.e. in total about two
e-foldings, or an expansion of the Universe by a factor of $7.4$. If we were more conservative and defined the
transition to go from $\Omega_\Lambda(a) = 0.01$ to $\Omega_\Lambda(a) = 0.99$ then the transition
takes about three e-foldings, corresponding to an expansion by a factor of $20$. 
I leave it to the reader to judge whether we should consider this very fine-tuned
compared to seven e-foldings (factor of $1100$) from the emission of the cosmic microwave background radiation (CMB) to today, or the roughly 
74 e-foldings between the Planck time and today.

These problems notwithstanding, it is still the case that the cosmological constant is in agreement
with all the data, and in addition it remains the best-motivated model. Is it really worth investigating
other possibilities? Apart from the basic truth that in order to test a model, we need to go beyond it, there is 
another reason why I think that the study of dark energy is worthwhile: according to the cosmological
standard model, the current period of accelerated expansion is not the first one. The standard model
postulates that in the very early
Universe, it was also accelerated expansion, called inflation, that created the seeds of the
structure that we see around us.
Was inflation due to a cosmological constant? It cannot have been a ``true'' cosmological constant,
since inflation ended. But can we test whether it was a kind of pseudo-static vacuum energy, or
something dynamical?

Inflationary dynamics is usually parameterized in terms of slow-roll parameters, with the first two
(in the Hamilton--Jacobi formalism, see e.g.~\cite{Liddle:2000cg}) defined as
\be
\epsilon_H = 2 \mpl \left(\frac{H'}{H}\right)^2 , \qquad \eta_H = 2 \mpl \frac{H''}{H} .
\ee
Here $'$ denotes a derivative with respect to the scalar field $\phi$ and
we used the reduced Planck mass $\mpl = 1/(8\pi G)$. With
$H' = \dot{H}/\dot{\phi}$ and with the help of the Friedmann equation (\ref{eq:friedmann})
and the conservation equation (\ref{eq:conservation}) as well as 
the expressions for energy density and pressure of a canonical scalar field (\ref{eq:scalar})
we find $\dot{\phi} = -2 \mpl H'$ and
\be
1+w = \frac{2}{3} \epsilon_H .
\ee
The equation of state during inflation is therefore directly given by
the first slow-roll parameter. To lowest order in slow-roll this is
also related to the tensor to scalar ratio by $r = 16\epsilon_H$. Without
any further work we can deduce that, since primordial gravitational waves 
have not been observed (yet?), there is no observational requirement for a 
deviation from $w=-1$ during inflation. The upper limit on $r$
from the seven-year Wilkinson Microwave Anisotropy Probe (WMAP)
data for a flat
$\Lambda$CDM model without running is about $0.36$ \cite{Komatsu:2010fb},  corresponding to
a maximum deviation from $w=-1$ of $0.015$, and adding additional data would
tighten this limit further.

So far it looks like inflation supports also an effective cosmological constant, with
$w=-1$ at the percent-level. However, we can do better by looking at the spectrum
of scalar perturbations:  the scalar spectral index $n_s$ is given to lowest order in slow roll by
\be
n_s -1 = 2 \eta_H - 4\epsilon_H .
\label{eq:ns}
\ee
A deviation of the
scalar spectral index from the Harrison-Zel'dovich (HZ) case ($n_s=1$) can be caused by the
second slow-roll parameter $\eta_H$. Thus even if at a given time $\epsilon_H \approx 0$, it is still possible
to obtain $n_s\neq 1$ through a non-zero $\eta_H$. But the rate of change of $w$ is also
linked to the first two slow-roll parameters,
\be
\frac{d\ln (1+w)}{dN} = \frac{d\ln \epsilon_H}{dN} = -2 (\eta_H-\epsilon_H)
\label{eq:wconst}
\ee
where $N=\ln a$ is again the number of $e$-foldings. Hence if Planck confirms the deviation
from HZ that WMAP sees at the 2.5$\sigma$ level, then not both $\epsilon_H$ and
$\eta_H$ can be zero. And if not both are zero, then either $w\neq-1$ or $w$ is changing
with time during inflation, or both. In either case, the result is not compatible with an effective cosmological constant
for the time period when the observable scales left the horizon, and we will be able to
conclude that {\em inflation was driven by a kind of early dynamic dark energy.} The Planck
measurement of $n_s$ may therefore lead to the first direct experimental evidence of the
existence of dark energy dynamics \cite{Ilic:2010zp}.

\subsection{Other possible explanations}  \label{sec:models}

If we accept that the data is indeed showing a surprising behavior of the Universe and is not just
plain wrong, and that we should look beyond a cosmological constant (at the very least in
order to test the $\Lambda$CDM model), what possible explanations are there? Basically, we
have to revisit the (sometimes implicit) assumptions that we have made and judge whether
they were reasonable and if there is a way to test them. We should not forget though that all
these alternative explanations also need to explain why the cosmological constant is
sufficiently small as to be irrelevant, when naively we would expect it to be far too large.

There are at least three fundamental assumptions that we usually make for the $\Lambda$CDM model:
\begin{enumerate}
\item The Universe contains the standard model particles, a dark, cold, matter-like component 
and a cosmological constant.
\item The large-scale evolution of the Universe is described well by General Relativity (GR).
\item The metric is well described by a Friedmann-Lema\^\i tre-Robertson-Walker (FLRW) metric with
small fluctuations.
\end{enumerate}

Let us start with the last assumption. Some theoretical justification is given by the Cosmological
(or Copernican) Principle that demands that there are no preferred observers in our Universe,
implying that it should be close to homogeneous and isotropic. Observationally, the CMB looks
very uniform and supports the Cosmological Principle. If  the Universe was exactly homogeneous
and isotropic then GR implies that the metric is of the FLRW form. A crucial question therefore is
whether the cosmological evolution of a state that is initially close to FLRW can
induce large changes from the expected evolution of the averaged metric. This scenario
is often called {\em backreaction} 
\cite{Buchert:1999er,Rasanen:2003fy,Rasanen:2006kp,Buchert:2007ik}, 
and since it is discussed elsewhere \cite{cras:clarkson2012} I will not consider it
further\footnote{I include here models that explicitly violate the Copernican Principle through the use
of a fundamentally inhomogeneous metric like LTB 
\cite{Moffat:1991qj,Tomita:2000jj,Celerier:1999hp,Biswas:2006ub,GarciaBellido:2008nz,Clarkson:2010uz} -- in my personal opinion, these models
are more likely to represent effective descriptions of the backreaction scenario \cite{Larena:2008be} 
rather than fundamental
models themselves, although large voids could potentially have been formed during inflation
\cite{Linde:1994gy}.}.
The only question I want to raise here is why in such a scenario the apparent evolution
of the Universe should be close to $\Lambda$CDM for all existing probes. It is not obvious
why the backreaction scenario should evolve towards a de-Sitter like state. 
So although backreaction models can potentially avoid some of the fine-tuning issues
inherent in $\Lambda$CDM, they seem subject to others that are not of a theoretical but of
an observational nature.

The other two assumptions are more closely related to the topic of this review, namely that
GR may be modified on large scales, and that there may be more in the Universe than the
particles that we know about (plus the dark matter). Here I will concentrate on the first
assumption, not least motivated by the discussion in the last section, namely that during
inflation we seem to observe a behavior that already requires extra ingredients beyond
those found in $\Lambda$CDM. But as
we will see, the formalism that we will put in place in the next section is equally able to deal with modifications
of GR, at least in an indirect way.

\section{A phenomenological description}  \label{sec:pheno}

\subsection{Overview}

Given the range of different explanations, it is useful to put in place a general and unified description
of possible observations. In this way, we can analyze the data without the need to assume a specific
model, and thus without imposing in advance limitations on what can be tested.

The best way to build such a framework is by reconstructing the metric perturbations. Here we 
will restrict ourselves to scalar perturbations on a flat background for simplicity, but the approach
can be generalized (and in general will need to be generalized once we consider non-linear perturbations
as then the scalar-vector-tensor split is no longer conserved). The metric
perturbations are good variables since they describe all observations that rely on gravity alone, specifically
{\em all} cosmological probes\footnote{I exclude here observations of non-gravitational physics like
those of spectral lines in quasars that probe the variation of fundamental constants, even though
such observations would provide a powerful way to probe the presence of light fields \cite{Parkinson:2003kf}.
See e.g.~\cite{Uzan:2010pm} for a review on varying constants and cosmology.}. 
Once the metric is given (as a function of space and time), we know that
the observations are now also fixed as the metric describes the motion of both light and of the test
particles (excluding astrophysical effects on small scales that we will neglect as we concentrate on observations
on large scales $k\lsim 0.1/$Mpc -- see sections \ref{sec:nonlinear} and \ref{sec:environment} for short remarks 
concerning non-linear effects on smaller scales).

The basic outline is then as follows: the line element that we will consider is
\be
ds^2 = a^2 \left[ -(1+2\psi) d\eta^2 + (1-2\phi) dx^2 \right] . \label{eq:metric}
\ee
Here $\eta$ is conformal time, and we will denote the physical time as $t$, with $dt = a d\eta$.
We therefore have one scale factor $a(t)$ and two metric potentials $\phi(x,t)$ and $\psi(x,t)$ that we 
need to know in order to fix the cosmological evolution of the Universe.

Through the Einstein equations
\be
G_{\mu\nu} = 8\pi G_N T_{\mu\nu} \label{eq:einstein}
\ee
 we can link the quantities in the metric to a physical description of the constituents of the Universe.
We can do this not only for a ``typical'' dark energy model where an extra component included in
the total energy-momentum tensor on the ``right hand side'' of the Einstein equations is responsible 
for the accelerated expansion, but also for models where the Einstein equations themselves are modified.
Even in that case we can use the observational fact that we are 3+1 dimensional beings to always
project the equations down to 3+1 dimensions, where we can use the (fundamental or induced)
metric to construct the Einstein tensor $G_{\mu\nu}$ and then build an effective dark
energy-momentum tensor by subtracting the $T_{\mu\nu}$ of the known components (baryons,
radiation and neutrinos):
\be
X_{\mu\nu} \equiv \mpl G_{\mu\nu} - T_{\mu\nu} \,
\ee
The new effective total energy momentum tensor on the right hand side is conserved (meaning 
$\nabla_\mu X^\mu_\nu = 0$) just because the Einstein tensor $G$ obeys the Bianchi identity $\nabla_\mu G^\mu_\nu = 0$
and because the EMT of the known particles satisfies $\nabla_\mu \tilde T^\mu_\nu = 0$.

How should we interpret this framework? In general, we need to interpret observations within
a theory. Without any theory, we do not know at all how to connect the observed distribution and
motion of e.g. galaxies with physics. We expect that the correct theory is similar to GR (and quite
possibly {\em is} GR). So we just start by assuming this to be the case, and derive the consequences
that arise from this assumption. If the resulting energy-momentum tensor looks suspicious, then
we will have to consider the possibility that GR may have to be modified. We will discuss in
section \ref{sec:modgrav} what such a suspicious signature of modified gravity could be.

\subsection{Background evolution}  \label{sec:background}

Let us start with a simple example of the framework that is well known and that has been used for
a long time: the so-called background evolution of the Universe, i.e. the evolution of the averaged
perfectly homogeneous and isotropic Universe. In addition, we take the Universe to be spatially flat for
simplicity. In that case, the 0-0 Einstein equation is the Friedmann equation,
\be
\left( \frac{\dot{a}}{a} \right)^2 = \frac{8\pi G_N}{3} \rho \, \label{eq:friedmann}
\ee
(with an overdot denoting a derivative with respect to physical time $t$), and the conservation equation becomes
\be
\dot{\rho} + 3 \left(\frac{\dot{a}}{a} \right) (\rho + p) = 0 . \label{eq:conservation}
\ee
In this case we have one free function that describes the geometry, $a$ or $H\equiv \dot{a}/a$, and two functions that
describe the contents of the Universe, $\rho$ and $p$. The Einstein equation provides a link
between the geometry and the contents. If we knew $\rho$ then we could use it to compute the
evolution of $a$. But in order to determine $\rho$ we need to know the pressure $p$. In this scheme it is therefore
the pressure $p$ that describes the physical nature of the ``stuff'' in the Universe. Conversely,
if we measure the evolution of the Universe, which corresponds to measuring $a(t)$, then we can
infer the evolution of $\rho$ and thus the evolution of $p$.

Usually we do not work with $p$ directly but instead we parameterize it with an ``equation of state'' parameter
$w$ through $p=w\rho$. Now the physical nature is given by $w$, and some simple examples are:
\be
w = \left\{ \begin{array}{rl}  1/3 & {\rm radiation \, / \, relativistic~particles} \\ 0 & {\rm dust~(matter \, / \, nonrelativistic~particles)} \\ -1 & {\rm cosmological~constant.} \end{array}  \right. 
\label{eq:standard_w}
\ee
If we therefore found that $p=\rho/3 \Leftrightarrow w = 1/3$ then we could conclude that the Universe
is filled with a gas of radiation.

What constraints on $w$ do we get from cosmological data? The basic approach is simple: we just
select a function $w(a)$ and compute $H(a)$ with the help of the Friedmann equation (\ref{eq:friedmann})
and the conservation equation (\ref{eq:conservation}). Based on $H(a)$ we can then compute predictions
for cosmological measurements that involve only the background geometry like luminosity and angular
diameter distances. We can then use for example a Markov-Chain Monte Carlo (MCMC) method to
evaluate the goodness of fit of the predictions with a likelihood function that is based on observations
of these distances (see section \ref{sec:quintobs} for an explanation of the MCMC method). 
In principle this is straightforward, but there are some technical questions. One
important question is how to choose the function $w(a)$. There are many different ways to parameterize
the equation of state parameter, for example as a function of redshift $z$ \cite{Huterer:2000mj,Maor:2000jy,Weller:2000pf}
or as a function of scale factor $a$ \cite{Chevallier:2000qy,Linder:2002et,Corasaniti:2002vg,Douspis:2006rs}. 
Although in principle both descriptions are equivalent, 
one has to be careful to keep the parameterization well behaved to $z\gsim 1000$ when using
high-redshift data, e.g.\ from the CMB \cite{Bassett:2002qu}. This is less of a problem with functions of the scale factor,
since it varies only in the interval $a\in [0,1]$, see \cite{Bassett:2004wz} for a discussion of different 
parameterizations and potential problems. An elegant approach is given by the Principal Component
Analysis (PCA), pioneered in a cosmological context by \cite{Huterer:2002hy}. In PCA the function
is expanded in a set of basis functions (for example in bins in $a$, but any functional basis will do)
and the covariance matrix of the coefficients is diagonalized. The eigenfunctions
then provide uncorrelated parameterizations of $w$, with the eigenvalues giving the precision with
which those functions of $w$ can be measured. The drawback of PCA is that it depends on the data
set used. Yet another possibility that has been investigated more recently is to use a
Gaussian process for modeling $w(z)$ \cite{Holsclaw:2010sk}.
For now we simply use a Taylor expansion of $w$ in $a$ and increase
the order until the goodness of fit no longer increases significantly.\footnote{An alternative approach to modeling the evolution of $w$ parameterizes instead the expansion rate $H$ or the dark energy
density $\rho_{\rm DE}$ \cite{Wang:2001ht,Tegmark:2001zc,Daly:2003iy}.}

\begin{figure}[htb]
\begin{center}
\includegraphics[width=3.5in]{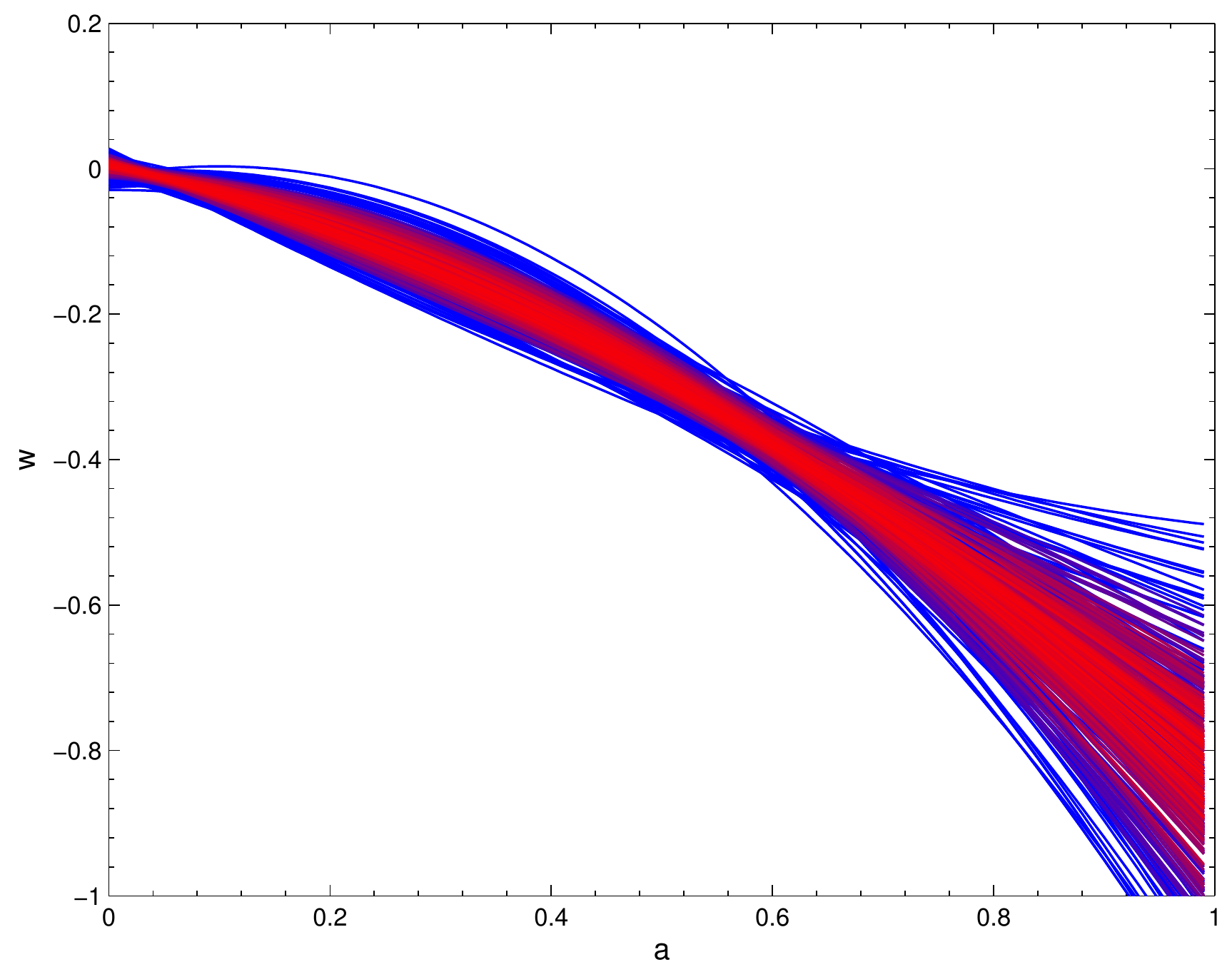}
\end{center}
\caption{An illustration of the form of the best fitting quadratic $w(a)$ curves. 
A sparse sampling of 400 chain elements,
  colour coded by likelihood with the red (lighter) shading the
  highest, is shown. (Figure from \cite{Kunz:2009yx})
\label{fig:total_w}}
\end{figure}

Another question is what data to 
use. Here one needs to be careful to avoid data sets that require the calculation of perturbations, the
necessary formalism for that is discussed in the following section. In \cite{Kunz:2009yx} we used
type-Ia supernova data (SN-Ia, we used the Union sample \cite{Kowalski:2008ez}), the peak position in the CMB
\cite{Wang:2007mza}
and Baryonic Acoustic Oscillations (BAO, \cite{Percival:2009xn}), as well as the SHOES measurement of $H_0$ \cite{Riess:2009pu}. We then
found the best-fit $\chi^2$ values listed in Table \ref{tab:modchi} (setting the spatial curvature to zero).
\begin{table}
\begin{center}
\begin{tabular}{llcc}
Model & equation of state& ~~Dark sector~~ & $\chi^2_{{\rm min}}$ \\
 &  & parameters & \\
\hline
Constant $w$ & $w=w_0$ & 1 & 391.3 \\
Linear (CPL) & $w = w_0 + (1-a) w_a$ & 2 & 312.1 \\
Quadratic & $w=w_0+w_1 (1-a)+ w_2 (1-a)^2$ & 3 & 309.8 \\
Cubic & $w=w_0+w_1 (1-a)+ w_2 (1-a)^2+w_3 (1-a)^3$& 4 & 309.6 \\
$\Lambda$CDM & $w=-(1-\Omega_m)/[\Omega_m a^{-3} + (1-\Omega_m)]$ & 1 & 311.9 \\
\hline
\end{tabular}
\end{center}
\caption{Number of parameters of the {\em total} dark sector equation of state (i.e.
dark matter and dark energy) and
  best-fit chi-squared for various parameterizations of $w$, for background-only
  data (SN-Ia, CMB peak position, BAO and $H_0$, see discussion
  in section \ref{sec:background}). The constant equation of state is ruled out
  while all others, including the $\Lambda$CDM equation of state, fit this data similarly well.}
\label{tab:modchi}
\end{table}
For comparison we also list $\Lambda$CDM which uses the single parameter $\Omega_m$.
For this data, a quadratic expansion of $w(a)$ appears sufficient, we plot in Fig.~\ref{fig:total_w}
a random sample of such quadratic curves to given an idea of the range of allowed
values of $w$. Although $\Lambda$CDM provides a slightly worse fit than the quadratic model,
it has fewer parameters and so passes this test. Only a constant total equation of state is
really ruled out -- we remind the reader that we consider here the total dark sector equation of state 
(i.e.  dark energy and dark matter together) since the Einstein equations relate the geometry
to the total energy-momentum tensor. We show in Fig.~\ref{fig:total_w} a random sample of the best fitting quadratic
$w(a)$ curves. At $a\lsim 10^{-4}$ the Universe is radiation dominated which implies $w=1/3$,
but the curves in the figure start later, during matter domination, and we can see that indeed
$p\approx0$ as expected for a Universe filled with pressureless dust.  But already
at relatively high redshift $w$ begins to decrease, and at $a\approx 0.5$ the expansion starts to accelerate
as $w$ drops below $-1/3$.
Today we have that $w\approx -0.8$, but with a large spread, mainly for two reasons: there is
little data at very low redshifts (where anyway the local dynamics starts to be important), and
 $w$ affects the distances only through a double integration which smoothes its impact strongly.

An additional subtlety here is that we are usually not interested in the total pressure $p$, but instead in the
pressure of the different constituents of the Universe. If for example the Universe was composed
of matter and a cosmological constant, then we would find a total $w$ somewhere between $0$ and $-1$,
see the last row of table \ref{tab:modchi} for the equation of $w$ in $\Lambda$CDM. Instead,
we would like to find two components, one with $w=0$ and a second one with $w=-1$. As we will
discuss in more detail in section \ref{sec:degen} this is in general not possible as the Einstein equations only
link the geometry with the total energy momentum tensor (EMT). For the time being we will assume that the
dark matter itself has been observed  so that we can separate the total EMT into several components,
\be
T_{\mu\nu}^{\rm (tot)} = T_{\mu\nu}^{\rm (radiation)} + T_{\mu\nu}^{\rm (baryons)} + T_{\mu\nu}^{\rm (DM)} + T_{\mu\nu}^{\rm (DE)} .
\ee
We put both electromagnetic radiation and neutrinos into the ``radiation'' EMT as we are not strongly
interested in these components here. DM denotes the non-baryonic dark matter, and DE whatever
it is that is responsible for the observed accelerated expansion of the Universe (including modifications
of gravity). We are then interested in the equation of state parameter $w_{\rm DE}$ that characterizes
the pressure of the DE component. We will discuss constraints on $w_{\rm DE}(a)$ in section \ref{sec:quintobs}
in the context of a canonical scalar field model for which the two contributions can be separated.

But even when we have measured $w_{\rm DE}(a)$ we will in general still not know what exactly the dark
energy is, except maybe if the result matches one of the simple cases given in Eq.~(\ref{eq:standard_w}).
In general there are many different possibilities that can lead to the same $w$. For example, a time-evolving
$w$ could be due to a scalar field as well as a modification of gravity. Even if, as is indeed the case, observations are consistent with
$w=-1$, we are still in trouble: we would like to interpret this result as a sign that the dark energy is
a cosmological constant, but for the reasons discussed earlier (and elsewhere in this volume), we think
that the cosmological constant is a problematic explanation for the dark energy.

Hence we would like to learn more about the nature of the dark energy. This is possible by exploiting
the extra information encoded in the
evolution of the perturbations in the Universe as we will discuss next. In addition, whenever we use
data that requires calculating the evolution of perturbations (like e.g. the CMB, the full galaxy distribution
or weak lensing) then we {\em must} provide a model that fixes the perturbation evolution. There is
no ``model independent way'' to just use $w$ in that case. We either need to pick an explicit physical
model or, as we will see in a moment, we need to choose precisely two extra functions to fix the
evolution of the scalar perturbations to linear order. Whenever we think that we can get away by
using only $w$, or maybe $w$ and the growth rate of matter perturbations, then we fix one or two
other quantities to some value (and in the case of multiple data sets possibly to several different, incompatible
values).

\subsection{Perturbation equations}  \label{sec:perturbations}

I will not enter into the details of cosmological perturbation theory since there are many
excellent texts on the topic, for example \cite{Kodama:1985bj,Durrer:1993db,Ma:1995ey,Hu:2004xd,Malik:2008im}.
Here we will work exclusively in the longitudinal (or conformal Newtonian) gauge. While the
background quantities like $\rho$, $p$ and $H$ are only functions of time or of the scale
factor, the perturbations are functions of both space and time. We will work in Fourier space as the linear perturbation
equations are particularly simple there since all $k$-modes evolve independently. The perturbations
themselves are random variables (so that in general we are really interested in the power
spectra), and due to statistical isotropy and homogeneity they are functions of $k=|{\bf k}|$ (and of time) only,
and independent of direction. We will further only consider scalar-type perturbations and assume that
vector and tensor-type perturbations can be neglected (which may not always be the case).

The conservation equations for a fluid with equation of state parameter
$w=p/\rho$ are then \cite{Ma:1995ey}
\bea
\delta' &=& 3(1+w) \phi' - \frac{V}{Ha^2} - 3 \frac{1}{a}\left(\frac{\dep}{\rho}-w \delta \right) \label{eq:delta} \\
V' &=& -(1-3w) \frac{V}{a}+ \frac{k^2}{H a^2} \frac{\dep}{\rho}+(1+w) \frac{k^2}{Ha^2} \psi 
	-(1+w)\frac{k^2}{Ha^2}\sigma  \label{eq:v}
\eea
where $\delta = \delta\!\rho/\rho$ is the density contrast, and we use
$V=i k_j T_0^j /\rho$ as the scalar velocity perturbation (the reason for this choice will become
clear in section \ref{sec:phantom}). A prime designates 
a derivative with respect to the scale factor $a$. The right-hand side of the
conservation equations also contains
the pressure perturbation $\delta p$ ($\sim$ the perturbation of the diagonal part of the
space-space part of the EMT) and the anisotropic stress $\sigma$ ($\sim$ the perturbation of the off-diagonal part of the
space-space part of the EMT). These four variables ($\delta\!\rho$, $V$, $\delta p$ and $\sigma$)
fully parameterize the scalar first-order perturbations of a general energy momentum tensor.
The conservation equations
need to be complemented by the Einstein equations. One of them gives the
$\phi$ potential in terms of the fluid content in a form reminiscent of the Newtonian
Poisson equation,
\be
k^2\phi =-4\pi G_N a^2 \sum_\alpha \rho_\alpha \left(\delta_\alpha+\frac{3aH}{k^2}V_\alpha \right) 
= -4\pi G_N a^2 \sum_\alpha \rho_\alpha \Delta_\alpha \, .
\label{eq:phi}
\ee
The sum over $\alpha$ on the right hand side runs over all fluids, and $\Delta$ is the comoving
density contrast (while $\delta$ is the density contrast in the longitudinal gauge).
The second potential is then related to the first through the anisotropic stresses present
in the fluids,
\be
k^2(\phi - \psi) = 12 \pi G_N a^2 \sum_\alpha (1+w_\alpha) \rho_\alpha \sigma_\alpha . \label{eq:sigma}
\ee

We notice that we have (at the perturbation level, i.e. without counting the background quantities)
two functions that determine the (scalar part of the) metric, $\phi$ and $\psi$, and four functions that describe the
behavior of a fluid and that define the (scalar part of the) energy-moment tensor, 
$\delta$, $\delta p$, $V$ and $\sigma$. Hence in total we have six functions, and four equations, so that
two of the functions cannot be determined. Given the structure of the perturbation equations where
the two conservation equations describe the evolution of $\delta$ and $V$, and the two Einstein equations
determine $\phi$ and $\psi$, it makes sense to consider $\delta p$ and $\sigma$ as the two free functions
that describe the type of fluid present in the Universe (although this is in principle a choice to be made).
This is very similar to the background case where the evolution of $\rho$ was determined by the
conservation equation and $H$ was given by the Einstein equation, while we considered $p$ (or $w$) 
as a free function describing the nature of the fluid.

Is it possible to construct a quantity like $w$ that is of order unity and easier to interpret than $p$,
to describe $\delta p$? For this purpose one often defines a rest-frame sound speed $\cs$,
defined through $\delta p = \cs \delta \rho$ for the comoving pressure and density perturbation. In terms of the quantities
in the conformal Newtonian gauge used here, this relation becomes
\be
\delta p = \cs \rho\delta+\frac{3aH\left(\cs-\ca\right)}{k^2}\rho V  \label{eq:soundspeed}
\ee
where $\cs$ is still the rest-frame sound speed, and where $\ca\equiv \dot{p}/\dot{\rho}$ is called
the adiabatic sound speed of the fluid. A fluid that has no internal degrees of freedom would have
$\ca=\cs$, but in general such fluids are not viable dark energy models unless they mimic
$\Lambda$CDM very closely due to the perturbation evolution \cite{Sandvik:2002jz,Bean:2003ae}. Models with
internal degrees of freedom like scalar field models (where both the field and its time derivative can be
seen as different degrees of freedom since the field obeys a second order equation of motion)
have in general $\cs\neq\ca$. We will see in section \ref{sec:quintpert}, where we discuss
the evolution of the perturbations in a generalized Quintessence model, that the role of the rest-frame sound
speed is really to describe pressure support, i.e.~it defines the existence of a sound horizon below
which the density perturbations do not grow. The sound speed here
is not necessarily related to the actual propagation velocity of perturbations, although the physics
of pressure support is usually related to the speed with which perturbations can adjust and thus
the sound speed coincides
with the propagation velocity for many models, most notably for the Quintessence and K-essence models that
we will discuss below. The sound speed does also not always provide a simpler description
than $\delta p$, for an explicit example where this is not the case see the Quintom model in section \ref{sec:phantom} below.

Finally, as already discussed in general terms in the introduction, the energy momentum tensor
considered here may be wholly or partially an effective, fictitious energy moment tensor. In this
case also the fluid quantities $\delta p$ and $\sigma$ will be effective quantities. (In these cases
as well we should not expect a simple $\cs$.) Nonetheless, we still can use the same formalism
as these functions provide a general and complete description of first order perturbation theory around a
FLRW metric. And as we will discuss in section \ref{sec:modgrav} especially the presence of a
non-zero anisotropic stress at late times (where contributions  from relativistic particles to the total
EMT are small) is a good indicator that we are dealing with a modified gravity model.

\subsection{Different descriptions}  \label{sec:description}

As we have seen, one way to describe the dark energy component is through the entries in the effective
energy moment tensor. But it is also possible to stay purely on the geometric side, and to parameterize just
the metric degrees of freedom. In this case, we describe the evolution of the Universe through the 
evolution of the scale factor, or more often through the Hubble parameter $H(a)$, as well as through the
two gravitational potentials $\phi(k,a)$ and $\psi(k,a)$. This is a description that it is quite close to the
observations. A closely related parameterization uses not the two metric potentials but instead two
linear combinations: $\Phi = \phi+\psi$ is the Weyl potential (its line of sight integral is the lensing
potential that describes the deflection of
relativistic particles or radiation due to gravity) as well as $\Pi = \phi - \psi$, the gravitational slip. Obviously
the pairs $\{\phi,\psi\}$ and $\{\Phi, \Pi\}$ provide an equivalent description of the perturbative
degrees of freedom. We should add here a health warning: there are different conventions in the literature for
the signs of the gravitational potentials as well as for which one is the time-time and which one the
space-space potential, so it is always a good idea to explicitly quote the metric used -- my choice is
given in Eq.~(\ref{eq:metric}), and I hope that I did not mix up conventions too often. Also the notation
for the anisotropic stresses is not uniform. Here I call $\sigma$ the fluid anisotropic stress in the energy momentum
tensor (with the appropriate pre-factors) and use $\Pi$ to denote the gravitational slip given by the difference of the two
gravitational potentials, which is the geometric quantity that is linked directly to the fluid anisotropic stresses,
see Eq.~(\ref{eq:sigma}).

But is the description in terms of the metric potentials  equivalent to the description in terms of the
fluid variables $\delta p$ and $\sigma$? In the direction fluid variables $\rightarrow$ metric potentials
it is easy to see that the answer is yes: given the fluid variables, we can use the Einstein equations
and the conservation equations to compute the evolution of the metric potentials. In the opposite
direction, it is less obvious. But since the Einstein equations (\ref{eq:einstein}) just identify, up to a
coupling constant, the Einstein tensor (which is a function of the metric d.o.f.) with the EMT, we can
project out the effective fluid quantities $\delta p$ and $\sigma$ also directly in the Einstein tensor
and see what functions of the metric we recover. Doing this \cite{Ballesteros:2011cm} we find that
the geometric anisotropic stress is indeed determined by $\Pi=\phi-\psi$ (effectively just Eq.~(\ref{eq:sigma})), and that
\be
\delta p_G = \frac{1}{4\pi G_N} \left[ \ddot{\phi} + H \left( 3 \dot{\phi} + \dot{\psi} \right)
-3 w_G H^2 \psi - \frac{1}{3} k^2 \Pi \right] \, .
\ee
Here we also used the total ``geometric'' equation of state parameter 
\be
w_G = -1 - \frac{2}{3} \frac{\dot{H}}{H^2} .
\ee
If we use this pressure perturbation together with the anisotropic stress inferred from the difference of 
the potentials and the background evolution given by $w_G$ then we find again the same
evolution of the metric potentials. The equivalence holds thus in both directions.

It is also possible to parameterize the geometric degrees of freedom through dimensionless variables that
link two perturbative quantities, similarly to the way $w_G$ links $p$ and $\rho$. One example
\cite{Amendola:2007rr} is given by the pair $\{Q,\eta\}$ defined through
\bea
k^2 \phi(k,a) &=& -4 \pi G_N a^2 Q(k,a)  \rho_m(a) \Delta_m(k,a) \label{eq:def_Q} \\
\psi(k,a) &=& \big[ 1+\eta(k,a) \big] \phi(k,a) \label{eq:def_eta}
\eea
In this way, $Q$ parameterizes both a deviation from the Poisson equation due to a modification of GR and
any extra contributions from perturbations beyond those of matter\footnote{Here we mean dark energy perturbations,
i.e. we implicitly assume that the parameterization is used only at late times. At early times the radiation perturbations
may be important, in which case they should probably be included explicitly.}. The latter is the case for
example if we have a Quintessence-like model (see section \ref{sec:quintessence}) for which the dark energy
perturbations are non-zero and will lead to $Q\neq 1$. The variable $Q$ can also model a variation of the
gravitational constant $G_N$.
Of course, in order to be general, both $Q$ and $\eta$ need to be functions of scale and time. This leads
to an additional complication, namely that a multiplication in Fourier space corresponds to a convolution
in real space, and vice versa. The conventional choice when limiting ourselves to first order perturbations
is to define $Q$ as a multiplicative factor in $k$ space where the equations are easier to deal with as
the modes decouple.

The choice of $Q$ and $\eta$ is not unique at all. The common theme is that we need to choose two
non-degenerate functions to parameterize the extra perturbations. One often defines a parameter
$\Sigma$ that is relevant for lensing through $\Sigma = Q(1+\eta/2)$ so that the lensing potential is given by
\be
k^2 \Phi = k^2 (\phi+\psi) = -8 \pi G a^2 \Sigma \rho_m \Delta_m .
\ee
Another often used parameter is $\mu = (1+\eta)Q$ which quantifies the impact of the dark energy perturbations
on a Poisson equation in $\psi$ in the same way as $Q$ for $\phi$. A compilation of different conventions
can for example be found in \cite{Daniel:2010ky}, see also \cite{Bean:2010zq,Ferreira:2010sz,Pogosian:2010tj} for other discussions.

For use in data analysis, some combinations may be better suited than others because they may
be more or less correlated. For example \cite{Zhao:2010dz,Daniel:2010yt} found that $\{\mu,\Sigma\}$ is a good choice for weak
lensing, CMB and large scale structure data. We have also left out the question of how to parameterize
the time and scale dependence of these functions. An early example is the Parameterized Post Friedmann
framework \cite{Hu:2007pj} (with an even earlier example, pre-dating dark energy, in \cite{Hu:1998kj}), 
while the papers listed above provide further possibilities. The most general and
flexible approach is probably the principal component analysis (PCA) approach already mentioned
in the context of parameterizing $w$. In PCA the two functions
are specified in a number of bins in $k$ and $a$, and the resulting covariance matrix between the bins is
then diagonalized to provide uncorrelated measurements of the resulting eigenfunctions.

\subsection{Schematic Measurements}  \label{sec:simpmeas}

Having discussed the freedom available, we also have to worry
whether the free functions can be measured at all in cosmology.
The aim here is not to discuss in detail the different observations. The goal of this short section
is merely to check whether it is possible to measure the gravitational potentials in principle.
Even this apparently straightforward question contains many different pitfalls. For example,
dark energy might couple directly to baryons and light, and possibly do so in different ways.
In this case we are going to reconstruct a metric that is not actually the metric we wanted to
reconstruct. On the other hand, we could have transformed the action so that at least e.g. baryons
are not coupled, and in general we assume that we did this. We will also assume that the
constraints from direct tests for fifth forces affecting photons and baryons are applicable also
on the scales of interest and limit such couplings to a level where they are not relevant.

If we are able to use the propagation of light and of baryons, then a possible high-level scheme
can proceed as follows: the propagation of light is governed by the perpendicular derivative
of the lensing potential $\phi+\psi$, so that with the help of weak lensing we are able to
constrain this combination of the potentials (this is also the reason why the $\Sigma$ parameter
is a good choice when using weak lensing data). Non-relativistic particles on the other hand
are accelerated by $\nabla\psi$ (the contribution from $\phi$ is suppressed by the particle
velocity). Observing the peculiar motion of objects on large scales, where they are not affected
by gas-physics, it is thus possible to constrain $\psi$. One possibility is to use redshift space
distortions on large scales for this purpose. In this way we can at least in principle obtain both
$\phi$ and $\psi$ from the observations. Once we have measured both potentials independently,
then we can e.g.~read off whether the anisotropic stress is zero or not, since $\sigma$
is directly related to the gravitational slip $\Pi=\phi-\psi$ through Eq.~(\ref{eq:sigma}).

We have not studied the measurements in any detail, but based on the above
discussion we can conclude that a combined large-scale structure and weak-lensing
survey can in principle constrain separately the two perturbation variables that characterize the dark
sector. This means that if we include these variables in the likelihood of the observations,
then we expect to obtain some constraints. How strong the constraints will be can be studied
for example with the help of the PCA approach mentioned in the last section.

Here we have not yet used galaxy clustering. The reason is that in modified gravity models
it is non-trivial to determine the bias between the clustering of galaxies and the clustering
of the dark matter (e.g.~\cite{Ballesteros:2011cm}). 
However, as we have seen, we do not need this measurement at least in principle.

\section{Dark energy models}  \label{sec:DE}

\subsection{Quintessence}  \label{sec:quintessence}

Let us consider how the basic canonical scalar field model looks like in the phenomenological
description discussed here. The action of the canonical scalar field (often called Quintessence 
or cosmon in a dark energy context \cite{Caldwell:1997ii,Wetterich:1987fm,Ratra:1987rm}) $\varphi$ is given by
\be
S = \int d^4\!x \sqrt{-g}  \big( -\frac{1}{2} \dd_\mu\varphi \, \dd^\mu \varphi - V(\varphi) \big) .
\label{eq:quintact}
\ee
From this action we can compute the evolution equation of $\varphi$ through variation with
respect to the field, and the energy momentum tensor through variation with respect to the
metric (if we added also the Einstein-Hilbert action then the variation wrt the metric would give
the Einstein equation). We usually split the field into a homogeneous background field (a kind
of condensate) and perturbations, $\varphi(k,t) = \bar \varphi(t) + \delta\varphi(k,t)$, and linearize
the equations in the latter.

For the background quantities, one obtains
\be
\ddot{\bar\varphi} + 3 H \dot{\bar\varphi} + \frac{dV}{d\varphi} = 0 \, , \quad
\rho = \frac{1}{2} \dot{\bar\varphi}^2 + V(\bar\varphi) \, , \quad
p = \frac{1}{2} \dot{\bar\varphi}^2 - V(\bar\varphi) \, . \label{eq:scalar}
\ee
We see the energy density and pressure depend on the potential and the evolution of
the field, so a Quintessence field can dynamically change $w$. But we also find
that $\rho+p = (1+w)\rho = \dot{\bar\varphi}^2 \geq 0$, implying that $w\geq -1$ as long
as $\rho > 0$. 

It is furthermore easy to show that the evolution equation for $\bar\varphi$ together
with the expressions for $\rho$ and $p$ is exactly the conservation equation
$\dot{\rho} + 3 H (\rho+p) = 0$. The analogous computation can be done with
the equations for the perturbations $\delta\varphi$, and one finds that they
correspond exactly to the fluid perturbation equations in section \ref{sec:perturbations} for a pressure perturbation
defined through (\ref{eq:soundspeed}) with $c_s^2=1$ and no anisotropic stress, $\sigma=0$ \cite{Hu:2004xd,Kunz:2006wc}.
To first order in perturbation theory, a canonical scalar field is therefore behaving just like a fluid
with speed of sound equal to the speed of light, and instead of evolving the Klein-Gordon equation
of the field, we can just use the fluid perturbation equations.

There are however many aspects that are not captured by the fluid description of Quintessence.
For example, the formalism gives no indication what equation of state $w$ to use. We would expect
in general the physics to specify a scalar field potential, which then allows to integrate the evolution
equation which in turn tells us what $w$ is. Of course this can be reversed: given a set of data, we
can use the fluid description to constrain $w$, which in turn places constraints on the allowed shape
of $V$ \cite{Huterer:1998qv,Saini:1999ba,Sahlen:2005zw}. Another important point that is missing from the fluid description
is the existence of fixed points and attractors in the evolution of $\varphi$, especially in the presence
of other fluids. Scalar field models with exponential potentials are for example able to adjust their
equation of state to follow the background evolution at a fixed fraction of the energy density. This
allows for a partial solution to the coincidence problem: The scalar field starts out with a fairly
arbitrary energy density early on, which then either decays rapidly or freezes until the scalar field
energy density roughly matches the radiation energy density. The scalar field then ``follows'' radiation
and latter matter. In this way it is natural that the scalar field today has an energy density comparable
to the one in matter. Unfortunately it has proven very difficult to engineer a natural exit from this
scaling phase, in general one needs to place a feature into the potential just at the right point.
A very nice discussion of the scalar field evolution described as a dynamical system with a discussion 
of scaling properties for different potentials can be found for example in \cite{Copeland:2006wr}.

\subsection{Generalizing Quintessence}  \label{sec:k_essence}

As we have seen, the fluid picture of Quintessence is determined through the parameter choices
$\{w,c_s^2=1,\sigma = 0\}$, i.e. $w$ is a free function, but the pressure perturbation and the anisotropic
stress are fixed. In this and the following section, we will see whether we can relax the two
conditions, and what the implications of such generalizations are.

A simple way to change the sound speed is to allow a different kinetic term in the action (\ref{eq:quintact}).
Writing $X \equiv -\frac{1}{2} \dd_\mu\varphi \, \dd^\mu \varphi$ the so-called K-essence action then becomes
\cite{Chiba:1999ka,ArmendarizPicon:2000dh}
\be
S = \int d^4\!x \sqrt{-g} K(X,\varphi)  ,
\label{eq:kact}
\ee
where $K(X)$ is a function of the standard kinetic term $X$, and we recover standard Quintessence
through $K(X,\varphi)=X-V(\varphi)$. As for Quintessence, there is a vast literature on K-essence models, here I only
summarize some standard results (see e.g.\ \cite{2010deto.book.....A}). As usual through the variation of the action
we can find the equation of motion and the energy momentum tensor. One finds that at the background
level the energy density and pressure are given by
\be
p = K \, , \qquad \rho = 2 X K_{,X} - K
\ee
where we dropped the dependence of $K$ on $X$ and $\varphi$ and where we used the
shortened notation $K_{,X} = dK/dX$. The equation of state is therefore
\be
w = \frac{p}{\rho} = \frac{K}{2 X K_{,X} - K}
\ee
which for Quintessence (where $K_{,X} = 1$) coincides with with the equation of state derived
from Eq.~(\ref{eq:scalar}). When studying the perturbations, one finds that also K-essence
models are described by a sound speed given by
\be
c_s^2 = \frac{p_{,X}}{\rho_{,X}} =  \frac{K_{,X}}{2 X K_{,XX} + K_{,X}}
\ee
and vanishing anisotropic stress, $\sigma=0$. K-essence models therefore generalize the Quintessence
models to $\{w,c_s^2,\sigma = 0\}$, where now $w$ and $c_s^2$ are free. The sound speed
in K-essence models is no longer equal to the speed of light. It can be larger or smaller than 1 and
also vary over time. We note that for the Quintessence case
we recover again the expected result $c_s^2=1$, while for the 
class of models with $K(X)=X^\alpha-V(\varphi)$ we have $c_s^2 = 1/(2\alpha-1)$, i.e. we can
choose a constant sound speed through the choice of  $\alpha$.

An important point (and a main motivation for introducing K-essence) that is not
apparent in a fluid formulation, is that in K-essence models the evolution of the homogeneous
``background'' field can be made to depend on the overall expansion rate of the Universe so that
the onset of matter dominated expansion triggers a transition to a later accelerated expansion stage
dominated by the K-essence field. That is, K-essence models can (for suitable choices of the kinetic
function) overcome the coincidence problem of the cosmological constant by linking the onset of
dark energy domination to the earlier onset of matter domination. As matter perturbations do not
grow during radiation domination, structure and therefore intelligent lifeforms similar to us
can also form only after the radiation-matter transition and so a coincidence between the emergence
of species observing the cosmos and the transition to dark energy dominated expansion is expected
in such models. However, a drawback of these models is that the speed of sound necessarily
becomes larger than unity in all models that can solve the coincidence problem \cite{Bonvin:2006vc}.
Whether such a superluminal propagation of information is problematic is a subject of debate
\cite{Bonvin:2007mw,Babichev:2007dw}: in general if information can be transmitted faster than light for all
observers, then it is possible to construct closed time-like curves and so to transmit information
into the past. This leads to obvious problems with causality. On other hand, if information travels only
faster than $c$ with respect to some observers, for example with respect to the rest-frame of the
K-essence fluids on which the perturbations propagate, then causality problems can be avoided
but Lorenz invariance is broken and we have introduced an effective aether through the presence
of the K-essence condensate.

\subsection{Phantom crossing}  \label{sec:phantom}

As seen in section \ref{sec:quintessence}, a canonical scalar field model cannot cross the ``phantom divide''
$w=-1$. This does not mean however that such a crossing is impossible: we will briefly discuss a
two-fluid example here, and another possibility is mentioned in section \ref{sec:degen}, namely that
coupled dark matter - dark energy models can lead to apparent phantom crossing. In addition, theories
where GR is modified can in general cross as well, since there is no ``real'' dark energy field present.
When analyzing data, and especially since the data indicates that $w \approx -1$,
it is important to not exclude the possibility $w<-1$ by construction. In this section we will review at
a purely classical level the behavior of the perturbations close to $w=-1$ and how to avoid problems
when $w$ crosses $-1$
(at the quantum level phantom fields would allow for spontaneous vacuum decay, but potentially
viable models could still be constructed, see e.g.~\cite{Carroll:2003st,Deffayet:2010qz}). An additional motivation
for this section is that it illustrates explicitly the limitations inherent in parametrizing the pressure perturbations
purely in terms of a rest-frame sound speed: although this is often a good choice that significantly simplifies
the description of the situation, this is not always the case.

When looking at the standard perturbation equations, as e.g. found in \cite{Ma:1995ey} where the
velocity perturbations $\theta$ are defined through $T_{0,i}^i \propto (\rho+p) \theta$, one finds terms like
\be
\dot{\theta} = -\frac{\dot{w}}{1+w} \theta + \ldots
\ee
The division by $1+w$ looks like a problem, as it would lead to a divergence in $\theta$ when trying
to cross $w=-1$. But since $\rho+p$ was factored out in the definition of $\theta$, the energy-momentum
tensor can stay finite even if $\theta$ diverges. Indeed, if the divergence of $T_{0i}$ does not go to
zero at crossing, then $\theta$ will necessarily diverge. But this is only an apparent problem and easily
cured by changing slightly the definition of the the velocity perturbations, and it is the reason why
we use here $V$ defined through $T_{0,i}^i \propto \rho V$. 

A more severe problem appears when we look at the pressure perturbation parameterized through
a sound speed in the rest-frame. The gauge transformation from the rest-frame to another frame
involves the adiabatic sound speed $c_a^2 = \dot{p}/\dot{\rho}$ which can also be written as
\be
c_a^2 = w - \frac{\dot{w}}{3 H a (1+w)} .
\ee
The adiabatic sound speed necessarily diverges at phantom crossing, except when crossing with
$\dot{w}=0$\footnote{A more detailed analysis shows that for $\cs=0$ the crossing is possible
even if $\dot{w}\neq0$ \cite{Kunz:2006wc} -- interestingly, this is the same condition as the one found in
\cite{Creminelli:2008wc} with the help of effective field theory, and an explicit example can be
found in \cite{Lim:2010yk}.}.
Since $\delta p$ appears directly in the EMT, its divergence implies a potential singularity
in the metric and is not acceptable on physical grounds. The reason for this divergence is
that the rest-frame of the dark energy is badly defined when $w=-1$, so we chose to fix the
pressure perturbation to a finite value in the one gauge that we should not have used. Seen in
this way, the problem becomes easy to remedy: we just have to keep $\delta p$ finite in another
gauge, for example in the longitudinal gauge. Then nothing at all happens at phantom crossing.
In practice, this is what we will do later on when using supernova and CMB data to put constraints
on $w$, allowing also for $w<-1$. More precisely, we regularize the adiabatic sound speed
by setting
\be
\tilde{c}_a^2 = w - \frac{\dot{w} (1+w)}{3 \HH [(1+w)^2+\lambda]} 
\label{eq:intca}
\ee
where $\lambda$ is a tunable parameter which determines how close to $w=-1$
the regularisation kicks in. A value of  $\lambda\approx 1/1000$
appears to work reasonably well \cite{Kunz:2006wc}.

\begin{figure}[htb]
\begin{center}
\includegraphics[width=3in]{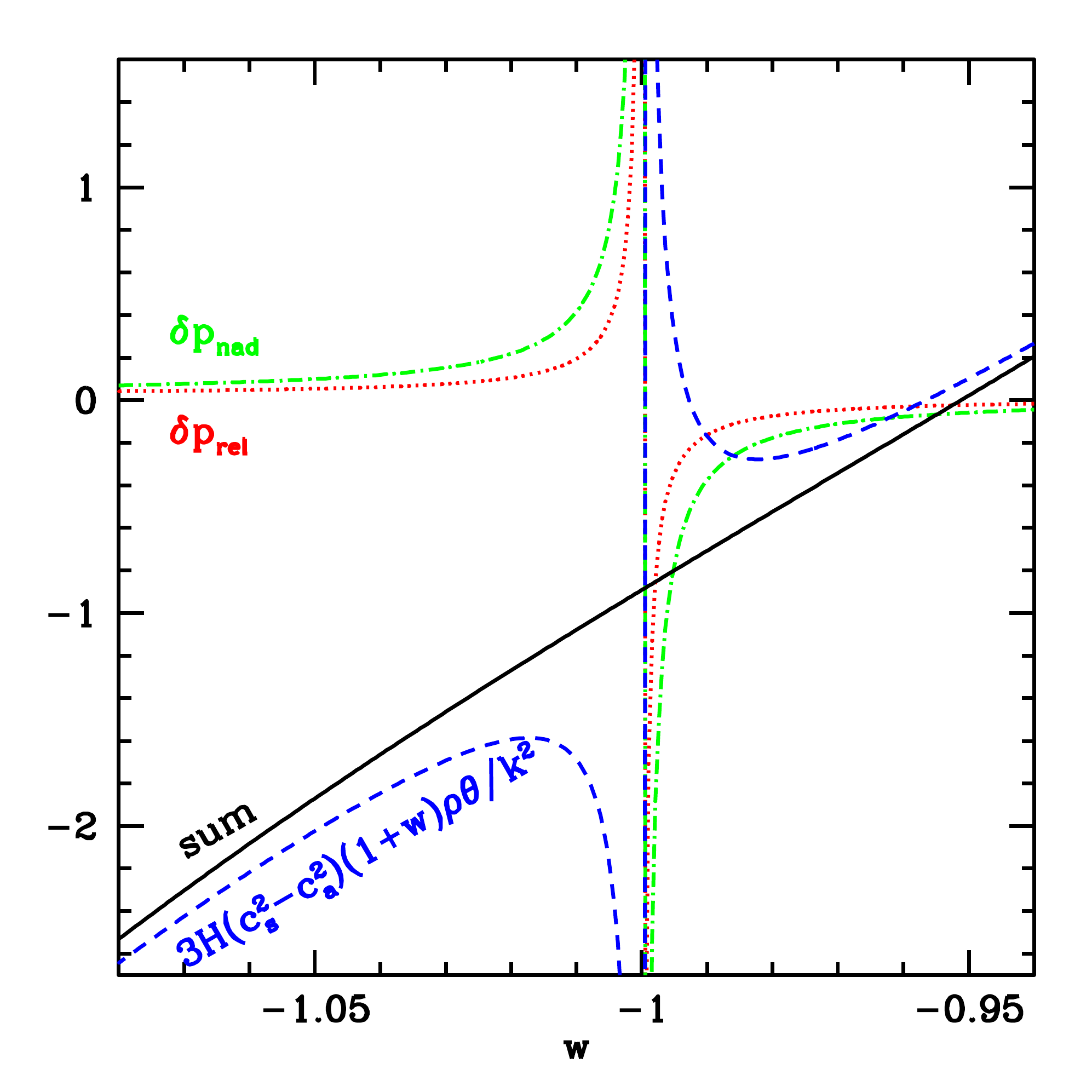}
\includegraphics[width=3in]{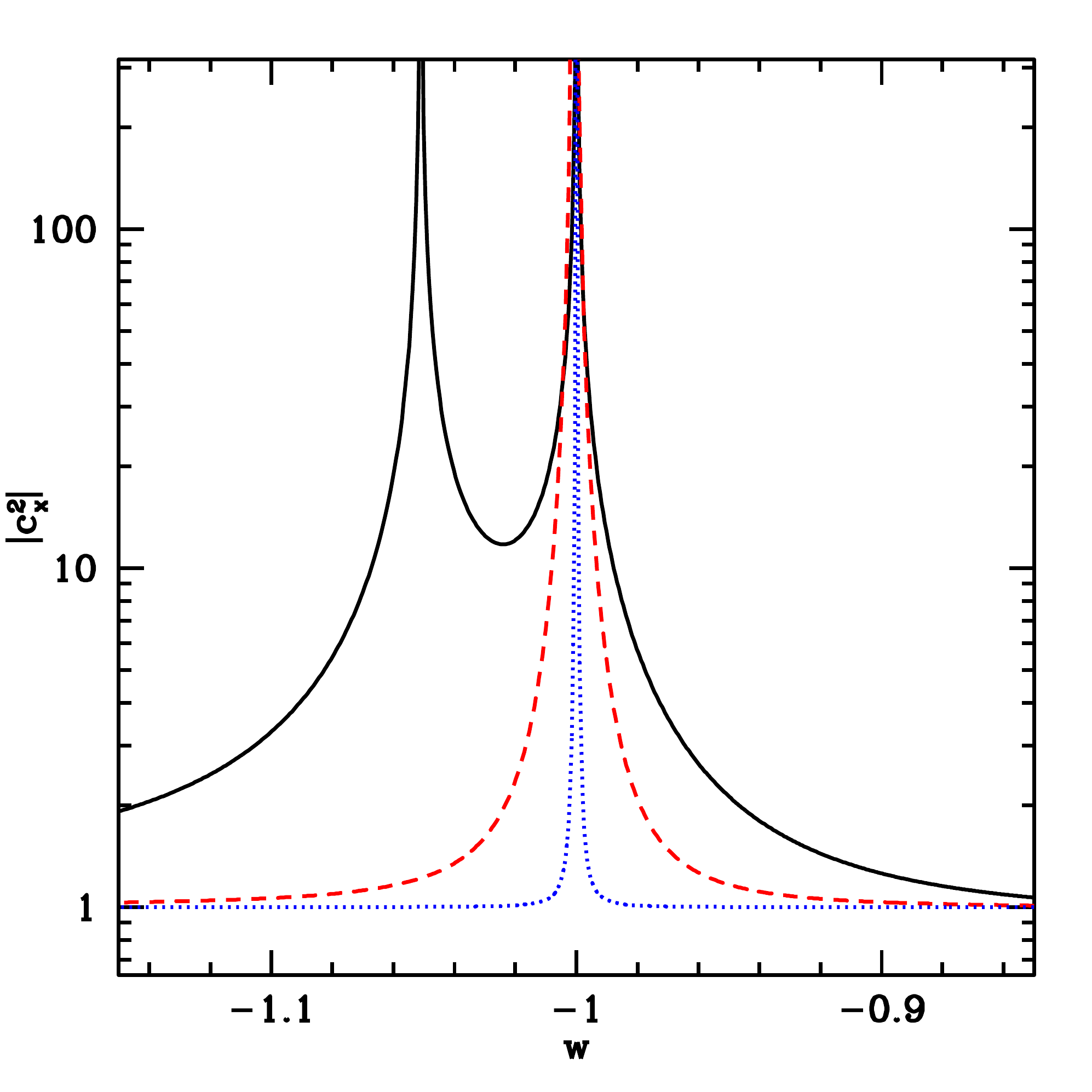}
\end{center}
\caption{{\em Left:} This figure shows the different divergent contributions to the pressure
perturbation, Eq.~(\ref{eq:dp_quintom}), multiplied by $10^9$. The relative
pressure perturbation is shown as red dotted line, the non-adiabatic pressure
perturbation as green dash-dotted line and the contribution from the gauge transformation
to the conformal Newtonian frame as blue dashed line. Each of the contributions diverges at
the phantom crossing, but their sum (shown as black solid line), and so $\dep$, stays finite.
{\em Right:} We plot the
apparent sound speed $c_x^2$ defined by Eq.~(\ref{eq:cx})
for three different wave vectors, $k=1/H_0$ (black solid
line), $k=10/H_0$ (red dashed line)
and $100/H_0$ (blue dotted line). 
Although the real sound speed is just $\cs=1$,
the apparent sound speed is scale dependent, diverges at $w=-1$ and can even
become negative. (Figure from \cite{Kunz:2006wc})
\label{fig:quintom}}
\end{figure}

Finally, it is instructive to have a quick look at the behavior of the perturbations in the
``Quintom'' model of dark energy \cite{Feng:2004ad,Hu:2004kh}. This model consists of two scalar fields,
one with a constant equation of state parameter $w_1>-1$ and another one with $w_2<-1$
(which necessitates changing the sign of the kinetic term, introducing a ghost degree of
freedom, but here we are only concerned with the classical behavior of the perturbations at the linear level).
As the ratio of the energy densities of the two fields scales as $\rho_2/\rho_1 = a^{-3(w_2-w_1)}$
and since $w_2-w_1<0$ we see that the second field becomes more important over time,
and so the total equation of state parameter will evolve from $w \approx w_1$ at early times
towards $w\approx w_2$ at late times, and will cross $w=-1$ somewhere in between. Yet
since we are just dealing with two scalar fields with constant $w$, we also know that nothing
strange will happen to the perturbations. In this system, we can compute in detail all the
perturbations and study their evolution \cite{Kunz:2006wc}. The total pressure perturbation can be written in terms
of the total effective quantities in the form
\be
\dep_\eff = \cseff \der_\eff+\dep_{\rel} +\dep_{\nad} 
	+ 3\HH\left(\cseff-\ca \right) \brho_\eff \frac{ V_\eff}{k^2}
\label{eq:dp_quintom}
\ee
where $\cseff = 1$ since this is the sound speed of both fields, $\dep_{\rel}$ is the contribution
from the relative density perturbation of the two fields (corresponding to a gauge invariant relative
entropy perturbation) and $\dep_{\nad}$ a non-adiabatic contribution from the relative motion of
the fields. We plot in the left panel of Fig.~\ref{fig:quintom} the evolution of these terms as the total
equation of state crosses $w=-1$. Each contribution individually diverges, but their sum remains
finite and well behaved. If one now tries to extract a total effective sound speed through the usual
formula
\be
\dep_\eff = c_x^2 \der_\eff +  3\HH\left(c_x^2-\ca \right) \brho_\eff 
\frac{V_\eff}{k^2} .
\label{eq:cx}
\ee
then all the terms get mixed up, and the resulting pseudo-sound speed $c_x^2$ is plotted
in the right-hand panel of Fig.~\ref{fig:quintom}. This pseudo-sound speed that is not connected
to any physical quantity has acquired a scale-dependence and additionally diverges at $w=-1$.

There are two main points to take away from this section, apart from the phantom-crossing discussion.
Firstly, although we have argued that the different parameterizations of the degrees of freedom available
in the metric and in the energy momentum tensor are in principle all equivalent, we have been provided here
with an explicit example that some parameterizations are nonetheless better than others, and that the
answer to the question of which one is best depends on the situation. Usually, the rest-frame sound
speed $\cs$ is a good parameter. It manages to capture the main characteristic of a canonical
scalar field in a single number ($\cs=1$) while a description in terms of $\delta p$ would look much more
complicated, and in addition the parameter itself describes a physical object
(the propagation velocity of the scalar field perturbations) of direct relevance to cosmology (as the
propagation speed gives rise to a sound horizon). However, in a situation where the direct physical
correspondence is lost, for example in the two-field Quintom model, the total ``rest-frame'' sound speed
takes a very complicated form, with scale dependence and divergences. Naively, one would probably
not have allowed such strange behavior, thinking it unphysical! For this reason, it appears important
to use several parameterizations and to test with forecasts whether we expect to see interesting effects
for the models where a given parameterization is simple.

But on the other hand, the phantom crossing example also shows how in some cases the
phenomenological approach allows a straightforward generalization of the behavior of the original
model. A canonical scalar field model would never cross $w=-1$. There is no choice of the potential
that can be made for which $w < -1$ (as long as the energy density is kept positive). If we just
restricted ourselves to reconstructing Quintessence potentials, we would never even have
realized that the possibility $w<-1$ exists. This is an example where the phenomenological
approach allows to probe the available degrees of freedom in a more model-independent way:
let us assume that the true model indeed has $w<-1$. Then, if all we knew were scalar fields,
we would presumably keep finding that $\Lambda$CDM is the best fit, even though there are
other models, that we did not think of, that would be better and could rule out $\Lambda$CDM.

\subsection{Perturbation evolution in generalized Quintessence models}  \label{sec:quintpert}

Let us take a closer look at the evolution of perturbations in Quintessence-type models with vanishing
anisotropic stress, but with an arbitrary constant sound speed $\cs$. In addition, we will assume that the equation
of state parameter $w$ is constant. This is strong assumption (and precludes for example the possibility
of early dark energy) but significantly simplifies the analysis in the fluid picture.
In this case the perturbation equations (\ref{eq:delta}) and (\ref{eq:v}) become
\bea
\delta' &=&  - \frac{V}{Ha^2}\left(1+\frac{9a^2 H^2\left(\cs-w\right)}{k^2}\right) \nonumber 
	-\frac{3}{a}\left(\cs-w\right)\delta+3\left(1+w\right)\phi' \label{deltap} \\
V' &=& -\left(1-3\cs\right) \frac{V}{a}+ \frac{k^2 \cs}{H a^2}\delta+(1+w) \frac{k^2}{Ha^2} \phi  \label{vp}
\eea
and we only have a single potential given by
\be
k^2\phi =-4\pi Ga^2 \sum_j \rho_j \left(\delta_j+\frac{3aH}{k^2}V_j\right) 
\label{eq:phiq}
\ee
since $\psi=\phi$ due to the vanishing anisotropic stress. Following \cite{Sapone:2009mb}
we also assume matter domination so that the background evolution is given by
\be
H^2 = H_0^2 \Omega_m a^{-3} .
\ee
For the matter dominated era, the (growing) solution of the matter perturbations ($w=\delta p = 0$) is well known,
\bea
\delta_m &=& \cc \left(a + 3 \frac{H_0^2 \Omega_m}{k^2}\right) 
= \cc a \left(1+3 \frac{H^2 a^2}{k^2} \right) \label{eq:deltam}\\
V_m &=& - \cc H_0 \sqrt{\Omega_m} a^{1/2}\\
k^2 \phi &=& -\frac{3}{2} \cc H_0^2 \Omega_m  \label{eq:phim}
\eea
which can be checked easily by inserting the solution into the perturbation equations. Since we
are working with linear perturbation equations, the overall scale is a free parameter, here called
$\delta_0$. It is in general a function of $k$ and is set by initial conditions that should include the
perturbation generation in the early Universe (inflation) and the subsequent evolution during radiation
domination. We also observe that the gravitational potential is constant in time.

We can now study the scalar field perturbations without the need to look at the actual field evolution,
since these equations are equivalent to the fluid equation with the appropriate value of the fluid
parameters. We first look at perturbations larger than the sound horizon,
$k \ll aH/c_s$. In this case, we neglect all 
terms containing the sound
speed in Eq.~(\ref{vp}), effectively setting $\cs=0$. The solution
for the velocity perturbation is (neglecting a decaying solution
$\propto 1/a$)
\be
V = - \cc (1+w) H_0 \sqrt{\Omega_m} a^{1/2} . \label{eq:vsub}
\ee
Up to the prefactor $(1+w)$ this is the same as for the matter velocity
perturbations. We find that this expression is valid on scales larger than
the sound horizon even if the sound speed is non-zero.

It is now straightforward to insert this solution for the dark energy velocity perturbation
into Eq.~(\ref{deltap}). Again setting $\cs=0$ we find the solution
\be
\delta = \cc (1+w) \left( \frac{a}{1-3 w} + \frac{3 H_0^2 \Omega_m}{k^2} \right) \label{eq:dsub}
\ee
where we neglected a term proportional to $a^{3 w}$ which is decaying
as long as $w$ is negative. Not surprisingly, also this solution becomes
equal to the one for matter perturbations for $w\rightarrow 0$. Relative
to the matter perturbations the dark energy perturbations are suppressed
by the factor $(1+w)$. This factor is necessarily always there, as the
gravitational potential terms contain it. It can be thought of as
modulating the strength of the coupling of the dark energy perturbations
to the perturbations in the metric. For $w=-1$ the dark energy perturbations
are completely decoupled (in the sense that they do not feel metric
perturbations -- but they can still produce them if the dark energy perturbations
are not zero).

The most important feature of the scalar field perturbations compared to the
matter perturbations is the existence of a sound horizon. Inside the light horizon,
the dark matter perturbations grow linearly with $a$ (until the perturbations
become non-linear). The dark energy perturbations on the other hand will
eventually encounter their sound horizon if $\cs>0$. Once inside the sound
horizon, they will stop growing. This means that the dark energy perturbation
spectrum is cut off on small scales.

To get a solution on small scales, $k\gg aH/c_s$, we start again with the equation
for the velocity perturbation. However, we expect the two terms with $k^2$
to cancel to a high degree to avoid large velocity perturbations, or in other
words
\be
\delta = -\frac{(1+w)\phi}{\cs} = 
\frac{3}{2} (1+w) \frac{H_0^2\Omega_m}{\cs k^2} \cc . \label{eq:dpsh}
\ee
As expected the dark energy perturbations stop growing and become
constant inside the sound horizon. 
The velocity perturbations are now given simply by using Eq.~(\ref{deltap})
and inserting Eq.~(\ref{eq:dpsh}):
\be
V = -3H a (\cs-w) \delta = - \frac{9}{2}(1+w)(\cs-w)\frac{H_0^3 \Omega_m^{3/2}}{\cs k^2} a^{-1/2} .
\label{eq:vsuper}
\ee
The extra term in brackets in Eq.~(\ref{deltap}) is not important for 
the scales of interest here.

\begin{figure}[htb]
\begin{center}
\includegraphics[width=3in]{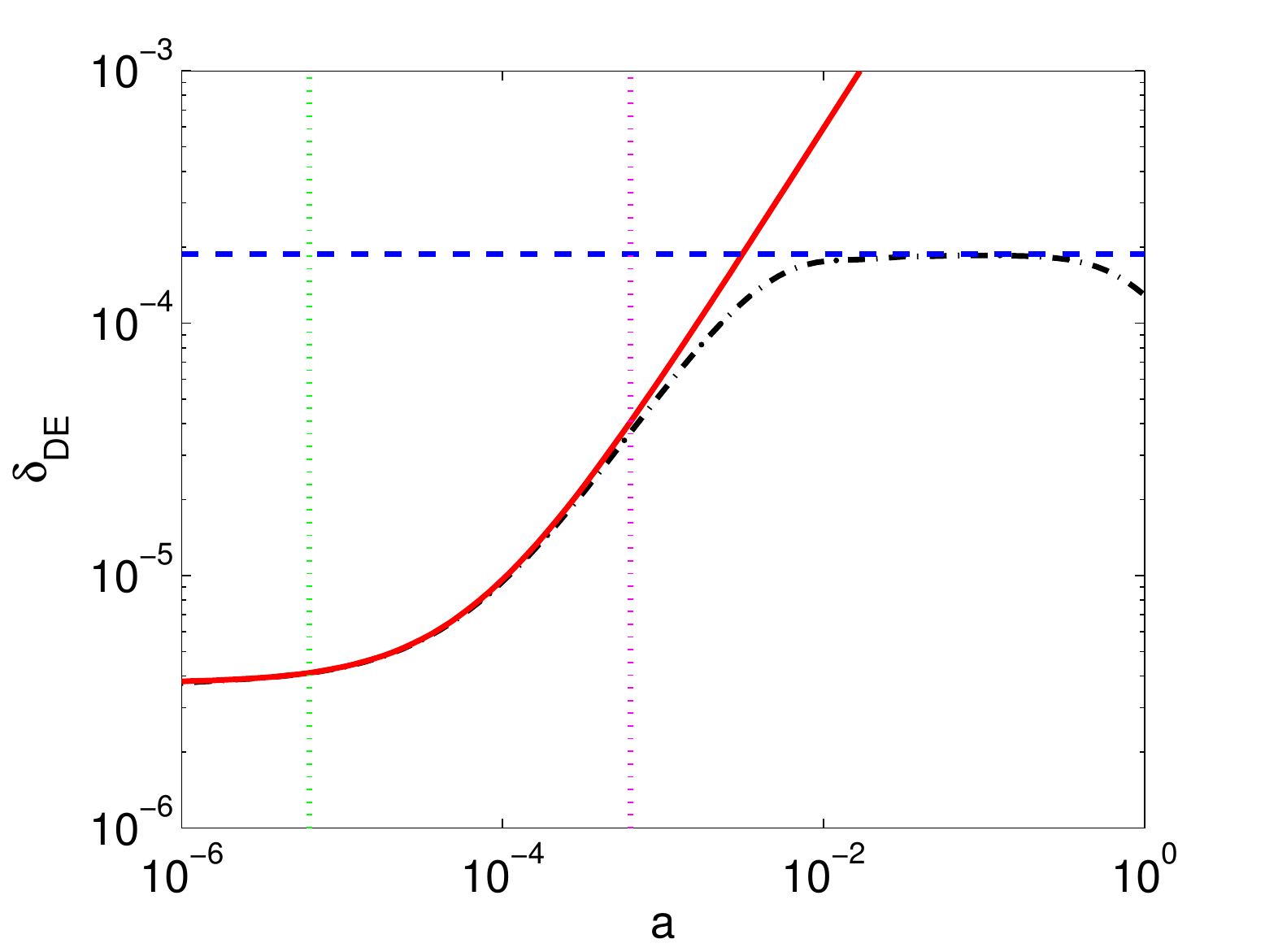}
\includegraphics[width=3in]{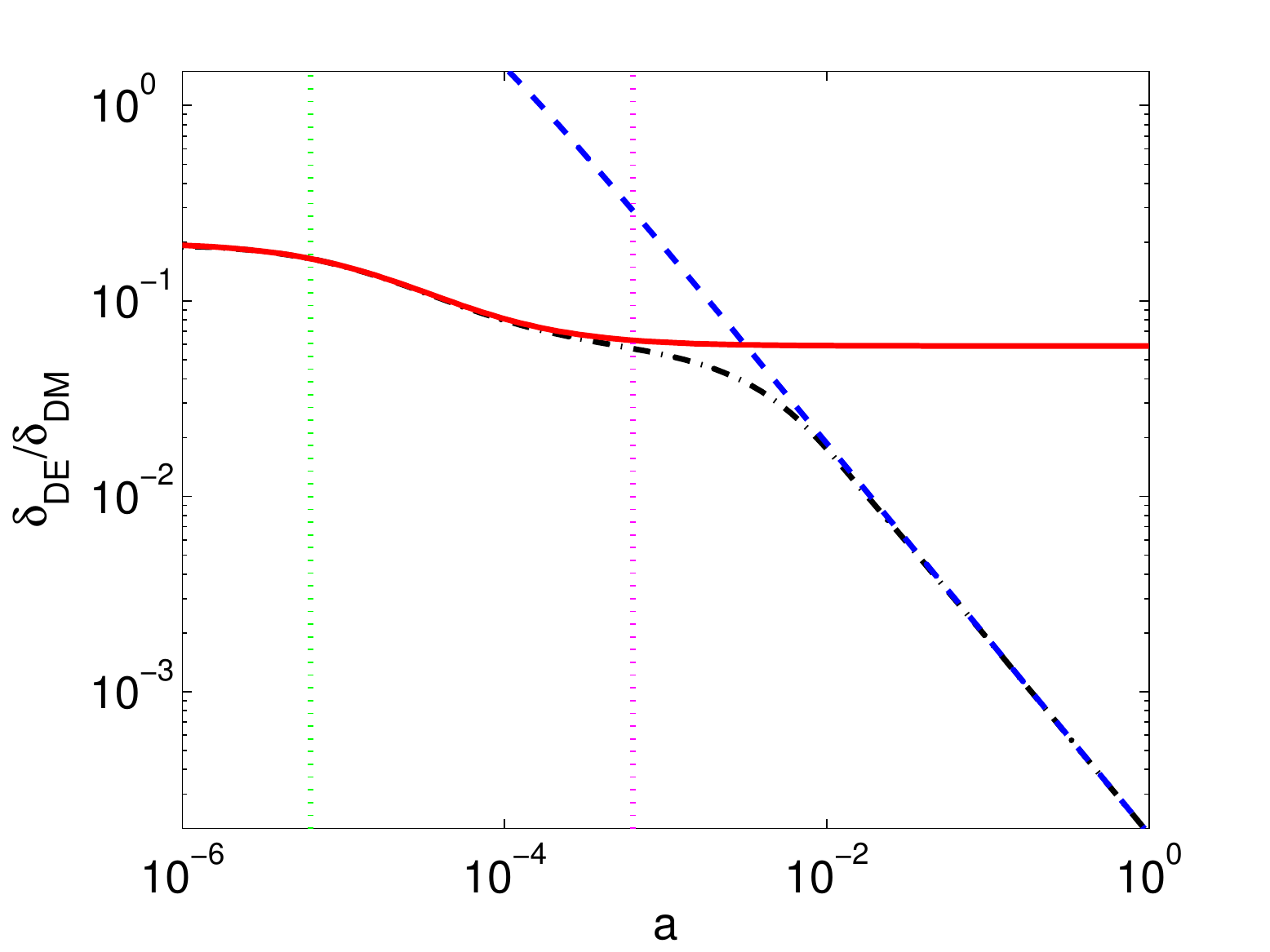}
\end{center}
\caption{{\em Left:} The figure shows the behavior of the variable $\delta_\de$
(in a universe without radiation). 
The black dot-dashed line is the numerical solution with $\cs=0.01$
and $w=-0.8$ for the mode $k=200 H_0$. 
The red solid line is the approximation on scales above the sound
horizon, Eq.~(\ref{eq:dsub}) and the blue dashed line is the approximation
to the scales below the sound horizon, Eq.~(\ref{eq:dpsh}). The two
vertical lines give the scale factor at which the mode enters the 
Hubble horizon (left line) and the sound horizon (right line). 
The numerical solution shows how the
perturbations decay at late times when matter domination ends, but radiation
was omitted from the numerical calculation to allow for a longer dynamic
range in $a$ to illustrate the different regimes.
{\em Right:} The ratio of dark energy to dark matter perturbations.
 For scales inside the dark energy sound horizon, the relative amplitude
decreases linearly with $a$. We can also see that the ratio of the perturbations
is described well by the fitting formula even when dark energy domination starts
to affect the perturbation evolution.
\label{fig:quintpert}}
\end{figure}

As the horizons grow over time, a fixed wave number $k$ will correspond
to a scale that is larger than the light horizon, $k<aH$, at early times,
and eventually it will enter the light horizon and later the sound horizon.
This makes it possible to illustrate the behavior of the perturbations
in the different regimes in a single figure: In the left panel of Fig.~\ref{fig:quintpert} we
plot the numerical solution for the dark energy density contrast
for $k=200 H_0$ as well as the expressions (\ref{eq:dsub}) and (\ref{eq:dpsh}).
It is easy to see how the perturbations start to grow inside the causal
horizon but how the growth stops when the sound horizon is encountered
and pressure support counteracts the gravitational collapse. 

In the right-hand panel of Fig.~\ref{fig:quintpert} we show the size of $\delta_\de$
relative to $\delta_m$. There are two effects: on large scales the dark energy
perturbations are suppressed by a factor proportional to $(1+w)$ relative to the
matter perturbations (and an additional factor $\sim 4$ inside the light horizon),
and as there is no sound horizon for perfectly cold dark matter, $\delta_m$ continues
to grow on small scales so that $\delta_\de/\delta_m \propto 1/a$ inside the dark energy
sound horizon. If we want to express the impact of the dark energy perturbations on
the gravitational potentials with the help of the $Q$ variable, as defined in
(\ref{eq:def_Q}), we should in principle
consider the rest-frame density perturbations which is a negligible change on small
scales and on large scales (and $w\approx -1$) just gives $\Delta_\de \approx \Delta_m (1+w)/4$.
Additionally we need to take into account the relative mean density, 
$\rho_\de/\rho_m = \Omega_\de/\Omega_m a^{-3w}$ for constant $w$. The combination
of these effects can then be expressed by the interpolating, approximate formula
\be
Q-1 = \frac{1-\Omega_{m}}{\Omega_{m}}\left(1+w\right)\frac{a^{-3w}}{1-3w+\frac{2}{3} \nu(a)^2}\label{eq:qtot}
\ee
where we introduced $\nu(a)^2 \equiv k^2\cs a/(H_{0}^{2}\Omega_{m})$ (the amount by
which a mode is inside the sound horizon)
and also assumed flatness so that $\Omega_\de = 1-\Omega_m$. This expression
is quite good on scales much larger and smaller than the sound horizon, while close to
the sound horizon it misses some transient effects. From this formula, we can see that today
($a=1$) the impact of the dark energy perturbation on large scales ($\nu\ll 1$) is of the
order of $3(1+w)/4$, while on small scales ($\nu\gg 1$) it behaves roughly like
$4 (1+w)/\nu^2$. On scales above the sound horizon therefore the perturbations
can contribute up to 10\% to $\phi$ given today's limits on $w$, but inside the sound
horizon the dark energy impact is strongly suppressed. Additionally, as we go back into
the past, the dark energy contribution to $\phi$ scales like $a^{-3w} \sim 1/a^3$, so at a
redshift of $z=1$ we get an additional suppression by a factor of 6 to 8.

The direct detection
of the perturbations is important because there is much more information on models
encoded in the perturbation variables than in the evolution of $w$. Measuring the
behavior of perturbations is akin to finding a fingerprint at a crime scene. By matching
the perturbation fingerprint against model predictions we may be able to understand
what the physics behind the accelerated expansion is. Unfortunately the smallness of
the perturbations in Quintessence and K-essence type models make it very difficult
to directly measure the perturbations. In \cite{Sapone:2010uy} we found that even large
future weak lensing and galaxy clustering observations can only hope to measure the sound speed (which we
use as a proxy to decide whether it is possible to detect the perturbations) if $\cs< 10^{-4}$,
a conclusion also supported by other studies \cite{dePutter:2010vy,Ballesteros:2010ks}.

Another point to take away from this section is how the use of the fluid equations and fluid
variables (rather than, say, the potential $V(\phi)$ of a Quintessence model or the exponent
$\alpha$ of a K-essence model with $K(X)=X^\alpha$) allowed us to study the behavior of the
perturbations in way that is more abstract from the fundamental model point of view, but that
emphasizes the physical evolution of the perturbations and allows to derive relatively simple
yet quite accurate formulae in order to study the expected observationally relevant effects.
The main effect for these models is the existence of a sound horizon within which the
perturbation growth is suppressed by pressure support. Because of this sound horizon, only
perturbations in sufficiently ``cold'' dark energy with $c_s^2\ll 1$ can be detected.

\subsection{Observational constraints on the Quintessence equation of state}  \label{sec:quintobs}

Already one of the original supernova papers included constraints on a constant
equation of state parameter $w$ for the dark energy \cite{Perlmutter:1998np},
an analysis that was extended in \cite{Perlmutter:1999jt}. Later papers like \cite{Saini:1999ba}
and \cite{Corasaniti:2001mf} allowed for an evolving $w(a)$, parameterizing either the
distance (for a fixed $\Omega_m$) or the Quintessence potential. These early analyses
only considered distance data and neglected the evolution of the dark energy
perturbations. For this reason they needed to either fix $\Omega_m$ or use a sufficiently
stiff parameterization to break the dark degeneracy that we will discuss in section \ref{sec:degen}.

Considering the evolution of the dark energy perturbations is not only important 
in order to use the additional information in probes like the CMB, but also because all
dark energy models apart from $\Lambda$ necessarily have perturbations, and
as demonstrated explicitly in \cite{Weller:2003hw} they are important even for
Quintessence models when using the full CMB data.
In \cite{Baccigalupi:2001aa,Bean:2001xy} the perturbations were taken
into account for a family of Quintessence potentials, and in \cite{Bassett:2002qu,Bassett:2002fe} we
parameterized directly $w(a)$ and reconstructed explicitly a Quintessence potential
for which the Klein-Gordon equation was solved. Since these papers used actual
Quintessence models, they always set $\cs=1$, while for example \cite{Bean:2003fb}
used directly the fluid equations and allowed for a free dark energy sound speed.

\begin{figure}[htb]
\begin{center}
\begin{tabular}{cc}
\includegraphics[width=3in]{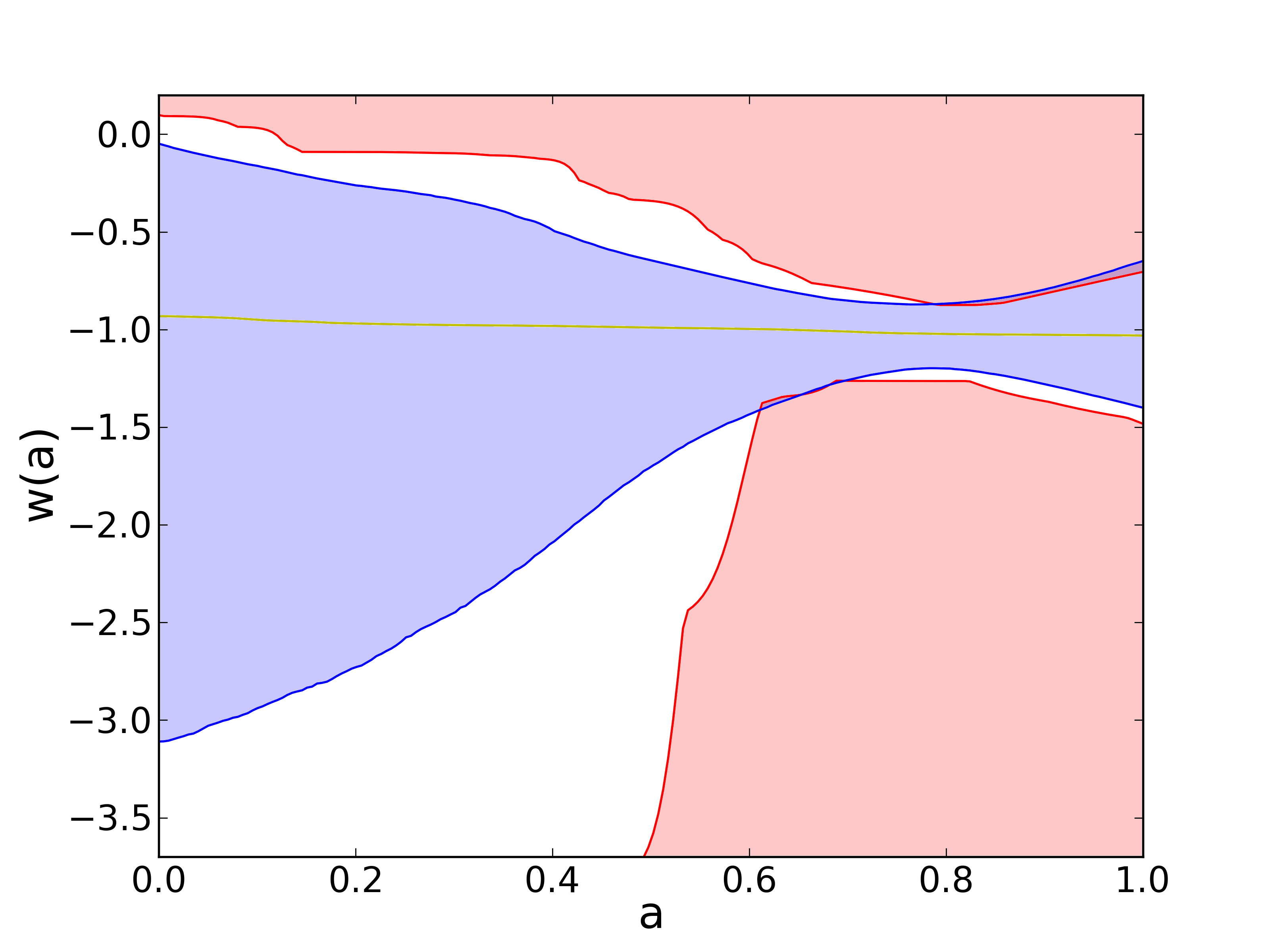}
\includegraphics[width=3in]{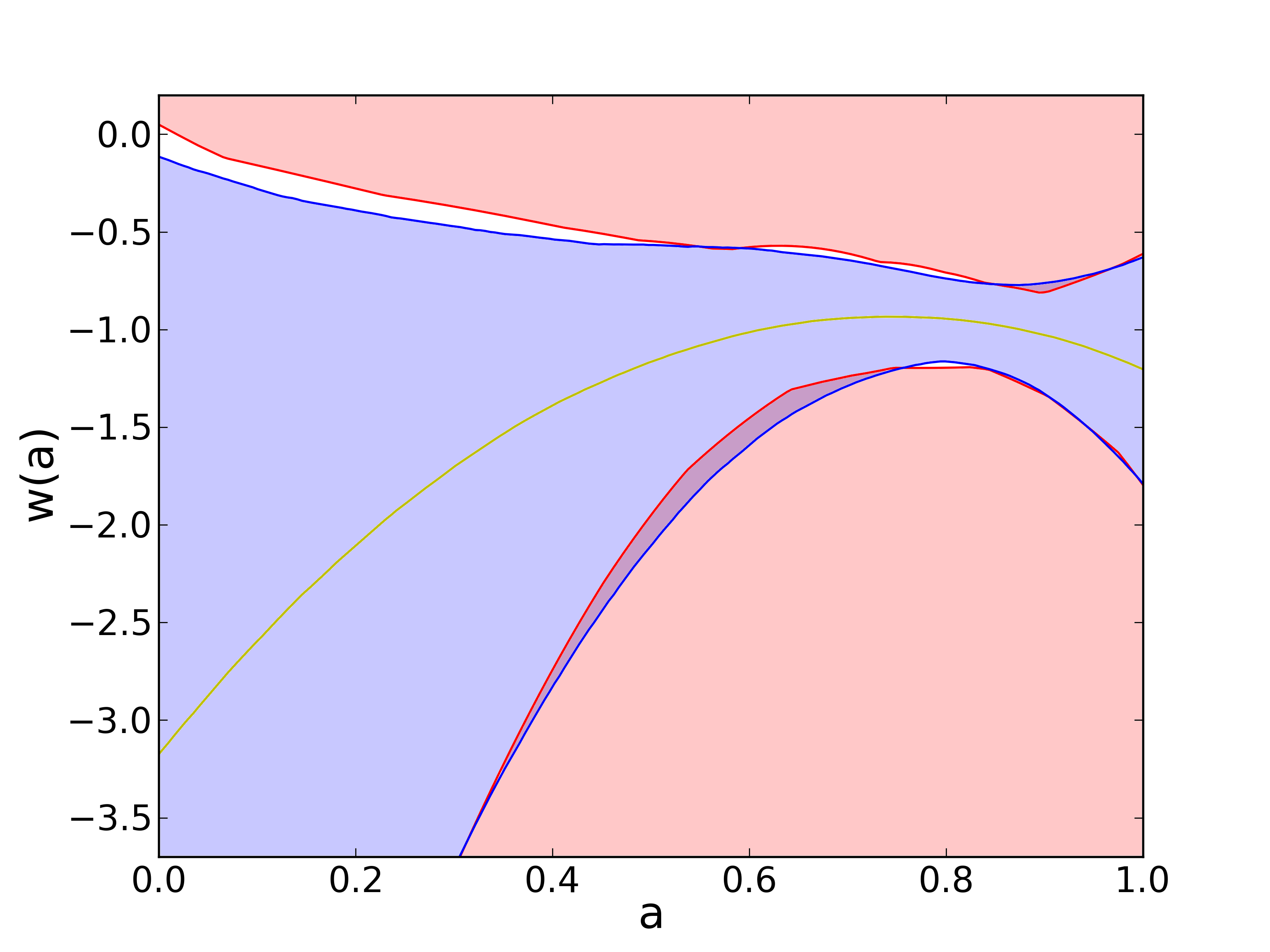}
\end{tabular}
\end{center}
\caption{Constraints on two parameterizations of $w(a)$ for a Quintessence model
$\{\cs=1$,$\sigma=0\}$ from WMAP CMB data and the Union type-Ia supernova data.
The blue (inner) shaded area corresponds to the location of 95\% of the accepted
$w(a)$'s, while any model in the red (outer) shaded area has an effective $\chi^2$
that is at least 4 worse than that of the best model. The left panel uses a kink parameterization
of $w$ \cite{Corasaniti:2004sz} while the right panel uses the same second order polynomial parameterization
as Fig.~\ref{fig:total_w}. Phantom crossing was modeled using the prescription of Eq.~(\ref{eq:intca}).
The constraints are best at $a\approx 0.8$ where we find that $w\approx -1 \pm 0.1$. At later
times the constraints are weaker due to the integrated nature of the data. A very early times,
$a \lesssim 0.4$, there is no strong lower bound on $w$ since very negative values make
the dark energy just vanish more rapidly in the past.
\label{fig:quint_w}}
\end{figure}

In this section we follow the approach of
\cite{Kunz:2003iz,Corasaniti:2004sz}, where the fluid perturbations (effectively (\ref{eq:delta}) and
(\ref{eq:v}) except that most Boltzmann codes work in the synchronous gauge rather than the
longitudinal gauge) were inserted into a modified
version of {\sc cmbfast} \cite{Seljak:1996is}. Differences are that we use a 
modified {\sc camb} \cite{Lewis:1999bs} instead of {\sc cmbfast} to compute the constraints
shown in Fig.\ \ref{fig:quint_w}, 
and that we include additionally the phantom crossing prescription defined through the regularized
adiabatic sound speed of Eq.~(\ref{eq:intca}).

Computing constraints on the equation of state parameter of the dark energy is a priori straightforward,
as already outlined in section \ref{sec:background}:
we pick a model $\MM$, here defined through a parameterization of $w(a)$ and a choice of $\delta p$ 
and $\sigma$ (we will call the set of parameters $\theta$, which is in general a vector of length $m$
for $m$ different parameters and includes also the cosmological parameters). 
We also choose some data $D$. For the purpose of the statistical analysis, the data
is encoded in a {\em likelihood}, which is the probability density of observing the data for a given model
and a given set of parameters, usually written as a function of the parameters 
$\LL(\theta) = P(D|\theta,\MM)$. To give a simple example, if our model were a Gaussian probability
distribution function (pdf) with unknown mean $\mu$ and known variance $\sigma^2$ and we tried to measure
$\mu$ (so that $\theta=\mu$) from a set of data $D=\{x_1,x_2,\ldots,x_n\}$ drawn independently
from the Gaussian pdf then the likelihood would be
\bea
\LL(\mu) &=& P(D|\mu,\sigma^2) = \prod_{i=1}^n \frac{1}{\sqrt{2\pi\sigma^2}} \exp\left\{ -\frac{1}{2}\frac{ (\mu-x_i)^2}{\sigma^2} \right\} \\
&=& \frac{1}{\left( 2 \pi \sigma^2 \right)^{n/2}} \exp\left\{ -\frac{1}{2} \sum_{i=1}^n \frac{ (\mu-x_i)^2}{\sigma^2} \right\}
\propto e^{-\chi^2/2} \mathrm{~for~} \chi^2 =  \sum_{i=1}^n \frac{ (\mu-x_i)^2}{\sigma^2} \, . \label{eq:gausslike}
\eea
Traditional (frequentist) statistics works directly with the likelihood, while Bayesian statistics is interested
in the probability of the parameters given the data, $P(\theta|D,\MM)$. With the help of Bayes theorem the two can
be linked,
\be
P(\theta|D,\MM) = \frac{P(D|\theta,\MM) P(\theta|\MM)}{P(D|\MM)} \, .
\ee
The second term in the numerator, $P(\theta|\MM)$, is called the prior since it corresponds to the prior knowledge
of how the parameter values are distributed, before the data is taken into account. The quantity in the denominator,
$P(D|\MM)$ is independent of the parameters and thus just an irrelevant proportionality constant when trying to
constrain the parameters. It is however an important quantity for model-comparison purposes in the Bayesian
framework (e.g.~\cite{Liddle:2004nh,Saini:2003wq,Trotta:2005ar,Kunz:2006mc}). 
How to choose the prior is not an entirely simple question. For a parameter like the
mean which can be at an arbitrary {\em location}, a natural choice is to choose $P(\mu|\MM)$ constant. On the
other hand if we wanted to estimate the variance $\sigma^2$ that is rather independent of {\em scale}, a prior
of the form $P(\sigma|\MM)\propto 1/\sigma$ is the usual choice. A detailed discussion of priors
(and of Bayesian statistics in general) can be found in the book by Jaynes \cite{jaynes}. One of the
standard works on classical statistics are the two volumes by Feller \cite{feller-vol-1,feller-vol-2}.

Once we have chosen a likelihood we need to compute confidence intervals, i.e. find regions in parameter
space that encompass a certain percentage (e.g.~95\%) of the probability. This requires effectively to compute
an integral in a potentially high-dimensional parameter space. High-dimensional integrals are generally
very difficult to compute numerically; for about a decade now the preferred solution in cosmology has been to use 
a Markov-Chain Monte Carlo (MCMC) method with Metropolis-Hastings acceptance criterion \cite{Metropolis:1953am}.
The algorithm is very simple:
\begin{enumerate}
\item Pick an initial point in parameter space, $\theta_0$, and evaluate the likelihood at that point, $\LL_0 = \LL(\theta_0)$.
\item Choose a new point $\theta_1$ so that the probability of picking $\theta_1$ when at $\theta_0$ is the same as the probability of picking $\theta_0$ when at $\theta_1$. (This condition can be relaxed by allowing for a slightly more complicated
acceptance criterion in step 4 below.)
\item Evaluate the likelihood at the new point, $\LL_1 = \LL(\theta_1)$.
\item Accept the step with probability $P=\min(1,\LL_1/\LL_0)$. If the new point is accepted, set $\theta_0 = \theta_1$ and
$\LL_0 = \LL_1$.
\item Record $\theta_0$ as a new element in the chain (even if the step was not accepted and $\theta_0$ has not changed!).
\item Go to step 2 and keep going.
\end{enumerate}
The usual way to choose a parameter vector $\theta_1$ in step 2 above is by picking a random vector $\Delta\theta$ from a Gaussian pdf with zero mean and a given, fixed covariance matrix, and
setting $\theta_1=\theta_0+\Delta\theta$. A good choice for the covariance matrix of the step-distribution is an approximation
to the parameter covariance matrix (formally, when allowing for an infinite number of MCMC steps, it does not  matter what step-distribution is chosen as long as it is symmetric and can reach all of parameter space, but in practice it is very important for the efficiency of the algorithm). One also needs to remove an initial non-stationary period of the MCMC evolution, the {\em burn-in},
and one needs to ensure that all the parameter space has been probed sufficiently, i.e. that the chain has {\em converged}.
See e.g.~\cite{Lewis:2002ah} for more details on MCMC methods in cosmology.

The output of the MCMC method is a so-called chain of parameter values that provide a random sample from the posterior
distribution of the parameters $\theta$. Since the density of points is proportional to the value of the likelihood, we can
marginalize (integrate out) parameters by just ignoring them. To get the marginalized pdf of a given parameter $\theta_i$, we can just look at the histogram of the values of that parameter in the chain, while for two-dimensional confidence contours one
needs to determine an area that contains e.g.~95\% of the probability. This is typically done by discretizing the relevant
2D parameter sub-space onto a grid and associating to each grid-square the number of points in the chain that lie there.
From the discretized 2D-histogram it is then straightforward to derive the desired contour.
A popular package for cosmological MCMC applications, with likelihoods for many data sets and tools for post-processing chains,
is {\sc CosmoMC}\footnote{Available from \url{http://cosmologist.info/cosmomc/}} \cite{Lewis:2002ah,cosmomc_notes}. A good discussion of MCMC methods and extensions (and generally a lot of statistics) can be found in the book by MacKay \cite{mackay2003information}\footnote{Also available on David MacKay's website, \url{http://www.inference.phy.cam.ac.uk/itprnn/book.html}}.
There are also other numerical methods to infer parameter constraints and compute model probabilities.
One approach that has become quite popular in cosmology over the last few years is nested 
sampling \cite{Skilling04,Bassett:2004wz,Mukherjee:2005wg,Feroz:2007kg}, another one is 
population-Monte Carlo \cite{Wraith:2009if}.

For the constraints shown in Fig.\ \ref{fig:quint_w} we use a likelihood that combines the Union 2 supernova 
compilation \cite{Amanullah:2010vv} and the 
WMAP 7-year data \cite{Larson:2010gs}. As models we choose two different parametrizations 
of the equation of state $w(a)$, the kink parameterization of \cite{Corasaniti:2004sz}
and a second-order polynomial form as used for Fig.\ \ref{fig:total_w}. We set $\cs=1$ and $\sigma=0$,
which corresponds to a canonical scalar field model of dark energy (Quintessence). 
To these parameters we have to add the usual cosmological parameters 
$\{H_0,\Omega_m, \Omega_b h^2, A_s, n_s, \tau \}$ as well as
possibly nuisance parameters required by the likelihood (none in our case since we marginalize
analytically over the absolute supernova luminosity). Here the parameters $H_0$ and $\Omega_m$
describe together with $w(a)$ the background evolution (we take the universe to be spatially flat) while the baryon
density $\Omega_b$, the reionisation optical depth $\tau$ as well as the amplitude of the primordial fluctuations $A_s$ and the
scalar spectral index $n_s$ are necessary for the CMB predictions.

After performing the MCMC we end up with a chain consisting of a large number (in our case about $10^5$) of accepted
parameter values. Each of these values encodes an evolution of $w(a)$. The figure then shows
for each value of $a$ the location of the central 95\% of $w(a)$ as the blue shaded region. 

Instead of
showing where most curves lie, we can also show the region in the $(w,a)$ plane where the dark energy 
evolution histories provide a ``bad fit'' to the data, relative to the best-fitting $w(a)$.
Let us assume that at a given $a$ the probability distribution for $w(a)$ looks roughly like
the Gaussian likelihood of Eq.\ (\ref{eq:gausslike}), with $\mu$ being $w(a)$. The
best fit (the maximum likelihood value) of  (\ref{eq:gausslike})
can easily be shown to lie at the arithmetic mean $\bar x$
with an uncertainty given by $\bar{\sigma}^2 = \sigma^2/n$
(see e.g.~chapter 24 of \cite{mackay2003information}), so that the distribution of the mean $\mu$
is again a normal distribution around $\bar x$ with variance $\bar\sigma^2$.
The integral from $\bar x-2\bar\sigma$ to $\bar x+2\bar \sigma$
(the so-called two-sigma interval) encompasses about 95.4\% of the probability for the location
of $\mu$ (and the one sigma interval, i.e.~the 
integral from $\bar x-\bar\sigma$ to $\bar x+\bar\sigma$, about 68.3\%). 
The quantity $\chi^2(\mu) = (\mu-\bar x)^2/\bar\sigma^2$ increases by 4 when
we move $\mu$ two sigma away from the best fit $\bar x$. 
We use this relationship and define an effective $\chi^2$ through $\chi^2_\mathrm{eff} = -2 \ln \LL$
which again for the simple case (\ref{eq:gausslike}) agrees with $\chi^2$ up to an irrelevant normalization
constant. We then plot in Fig.\ \ref{fig:quint_w} also a red-shaded region showing
 the location of $w(a)$ for which $\chi^2_\mathrm{eff}$ is larger by at least 4
than the smallest $\chi^2_\mathrm{eff}$. This is then expected to delineate roughly the $2 \bar\sigma$
boundary for $w$ at a given $a$ and thus should approximately coincide with the boundary of the blue
region, but shows the region where the badly fitting $w(a)$ lie.

We notice that the best constraints occur at $a\approx 0.8$ where we find $w\approx -1.0 \pm 0.1$
($1 \sigma$ errors). The constraints are compatible with $w=-1$ at all redshifts, the prediction for a cosmological
constant. At higher redshift (smaller $a$) there is no lower bound for $w$ since the dark energy
is subdominant in the past, and for very low values of $w$ it merely disappears faster as $a \rightarrow 0$.
We also notice that while the red ``exclusion'' region is similar for the two parameterizations, the
blue region for the kink model is narrower. This is an example of the prior imposed by a parameterization
or model for $w$ on the results. This is not necessarily undesirable: the distance data is linked to
$w$ through a double integration, so that the resulting constraints on the equation of state are
strongly smoothed. If we would for example use higher and higher order polynomials then we would
find less and less constraints as the resulting $w(a)$ would oscillate more and more around $-1$.
For this reason it is necessary to impose some constraint on the allowed form of $w$. For an
example on how to do this in a Bayesian way with a maximum entropy prior, see \cite{Zunckel:2007jm}.

\subsection{Modified Gravity models and anisotropic stresses}  \label{sec:modgrav}

So far we have been looking at models with fairly general pressure perturbations but still vanishing
effective anisotropic stress. Is there a class of models for which it is natural to have $\sigma\neq 0$
in their effective energy momentum tensor, so that  $\Pi= \phi-\psi \neq 0$ even when the contribution from
relativistic particles is negligible? Indeed,
there is: Let us look at a quite general scalar-tensor action, now including gravity and matter,
\be
S = \int d^4\!x \sqrt{-g} \Big[ \frac{1}{2} \big( 1+f(\varphi) \big) R + K(X) - V(\varphi) + \mathcal{L}_m \Big] .
\label{eq:stact}
\ee
Here we have chosen a frame in which the matter is minimally coupled, so that it follows the geodesics
of $g_{\mu\nu}$, since this is the frame what we would generally reconstruct from observations of weak
lensing and the motion of galaxies \cite{Amendola:2012zz}. We can again
compute the Einstein equations by varying the action with respect to the metric, and arrange it in the
Einstein form (\ref{eq:einstein}). In this case we find a total, effective dark energy EMT with
$\Pi \propto f'(\varphi)$ which does in general not vanish. This action therefore gives an explicit example
of a model which has all the possible degrees of freedom that can be recovered from cosmological
measurements\footnote{See also \cite{Brax:2011aw} where an action with a conformal coupling
to the matter field is proposed as a generic way to parameterize modified-gravity models}.

The scalar-tensor model is not the only example. Let us go through several typical ``modified
gravity'' models (see e.g.~\cite{Tsujikawa:2010zza} for a recent review). 
During this exercise we will also consider the question whether it is possible to
revert to the case $\Pi = 0$. For the scalar-tensor model it is easy to see that this requires
$f(\varphi)$ to be constant. In other words, requiring the absence of anisotropic
stress forces us to go to the GR limit of the theory. A class of models closely related to the scalar-tensor
type is the is the $f(R)$ kind of models, with the gravitational part of the action given by
\be
S_g  \sim \int d^4\!x \sqrt{-g}  f(R)  
\label{eq:fRact}
\ee
for an arbitrary function $f$. In this case, the effective anisotropic stress is found to be
$\Pi \propto f''(R)$. A vanishing anisotropic stress then is only possible if $f(R) = R + \Lambda$
for a constant $\Lambda$, again just the GR action\footnote{In principle the anisotropic
stress also vanishes if $\delta R=0$ but this is a complicated condition on the evolution
of the gravitational potentials that appears unnatural since it would require a very peculiar
matter contribution to be compatible with the Einstein equations, and is at any rate not in 
agreement with observations.}. The problem can also be seen from a different angle by
noticing that $f(R)$ models contain an effective scalar degree of freedom with a mass linked
to the anisotropic stress (in the quasistatic limit) through $\Pi \propto 1/m^2$.
Turning off the anisotropic stress requires therefore to make this
``scalaron'' very massive and so effectively suppressing it, forcing the theory to revert to GR.

Another widely studied modified-gravity model is the so-called DGP (Dvali-Gabadadze-Porrati \cite{Dvali:2000hr})
model which is based on a 4D brane embedded in a 5D bulk. The gravitational action is taken to be
the 5D Einstein-Hilbert action together with an induced 4D Einstein-Hilbert action confined to the
brane. The relative strength of the two contributions is given by the crossover scale $r_c = M_4^2/M_5^3$.
The effective anisotropic stress of the DGP model is  proportional
to $1/\beta$ where $\beta = 1+ 2 r_c H w_{\rm DE}$ (see (\ref{eq:dgp_eta}) below).  Setting the anisotropic stress to zero requires
sending $\beta$ to infinity, which in turn only happens for $r_c \rightarrow \infty$, or $M_5 \rightarrow 0$.
But in this limit we have turned off the 5-dimensional nature of DGP gravity and are left with only the
usual 4D Einstein-Hilbert action, and thus standard GR.

With two extra degrees of freedom on the other hand, it should be possible to balance them
against each other and so turn off the anisotropic stress \cite{Saltas:2010tt}. An example is afforded by the $f(R,G)$
type models \cite{DeFelice:2009ak} with action
\be
S_g  \sim \int d^4\!x \sqrt{-g}  f(R,G) 
\ee
where $G=R^2 - 4 R^{\mu\nu}R_{\mu\nu} + R^{\mu\nu\alpha\beta}R_{\mu\nu\alpha\beta}  $ is the 
Gauss-Bonnet term, a topological invariant in 4D (i.e. it only contributes a boundary term in the
action integral)\footnote{We use the rather complicated looking combination $G$ since the extra
combinations of the curvature tensor will in general lead to higher order equations of motion. This in 
turn leads to problems because of the Ostrogradski instability \cite{Woodard:2006nt}. This problem is
avoided by using $R$ and $G$.}. 
Indeed it is now possible to build models that have no anisotropic
stress, but in general the condition $\Pi=0$ depends on the background, i.e.~a model has no
anisotropic stress during e.g.~matter domination but $\Pi\neq0$ when accelerated expansion sets in. It should in principle be possible to create functions
that both lead to the right sequence of evolutionary stages (radiation domination, then matter domination
and finally accelerated expansion) and that retains $\Pi=0$, but it would be a quite complicated and fine-tuned endeavor, and it does not
appear as if there was an easy way to link the anisotropic stress to the evolution, rather the contrary.

As an example, we fix the background evolution to be de Sitter. In that case we find that the
condition $\Pi=0$ is equivalent to
\be
f_{,RR} + 8 H_0^2 f_{,RG} + 16 H_0^4 f_{,GG} = 0 .
\ee
The general solution of this equation is given by models of the form
\be
f(R,G) = f_1\left( R - \frac{G}{4 M^2} \right) + R f_2 \left( R - \frac{G}{4 M^2} \right)  \label{eq:frg}
\ee
for $M = H_0$ and $f_1, f_2$ two arbitrary functions. 
There are a few problems with this solution: Firstly one needs to end up in a de Sitter state with exactly
the right expansion rate, given by the mass parameter in the original model, to have $\Pi=0$. And secondly,
we found in \cite{Saltas:2010tt} that the evolution of $H$ is unstable close to the de Sitter point for models
of the type (\ref{eq:frg}). In addition, also in
these models, the mass of a scalaron degree of freedom has to diverge in order to force the anisotropic
stress to vanish. This does not constitute a proof that it is impossible to have zero anisotropic stress
for viable and non-trivial $f(R,G)$ models, but it is nonetheless surprising how difficult it is to set up
such a scenario.

It may be instructive to also spend a few lines investigation the typical size of the anisotropic stress
present in the DGP model. We choose this MG model since the $f(R)$ and scalar-tensor models can
always be continuously extended to GR, and their free functional degree of freedom makes it difficult
to consider `the' typical model of their class. For DGP we only have a single free number to choose,
the cross-over radius $r_c$. The background expansion (assuming flatness) in DGP is given by (e.g.~\cite{Maartens:2006yt})
\be
H^2 - \frac{H}{r_c} = \frac{8\pi G_N}{3} \rho_m .
\ee
We will consider the second term on the left, $H/r_c$, as providing the dark energy contribution to
the expansion rate and so consider it as being effectively on the right hand side. If the Hubble parameter
today is given by $H_0$ and the relative matter density today is taken to be $\Omega_m$, then 
we want for consistency $H_0/r_c = H_0^2 (1-\Omega_m)$ or $r_c = 1/[H_0 (1-\Omega_m)]$. As
naively expected, we see that the crossover scale of DGP needs to be of the size of the horizon scale
today, which reveals the required fine-tuning in this model. Without going into a great deal more detail
(see e.g.~\cite{Lue:2004rj,Koyama:2005kd}), one finds for the perturbations that 
\bea
k^2 \phi &=& -4 \pi G_N a^2 \left( 1 - \frac{1}{3 \beta}  \right) \rho_m \Delta_m \label{eq:dgp_phi} \\
k^2 \psi &=& -4 \pi G_N a^2 \left( 1 + \frac{1}{3 \beta}  \right) \rho_m \Delta_m \label{eq:dgp_psi}
\eea
for $\beta \equiv 1-2 r_c H [1+\dot{H}/(3 H^2)] = 1+ 2 r_c H w_{\rm DE}$,
and the matter perturbation evolution proceeds as usual (i.e.~as given by Eqs.~(\ref{eq:delta}) and (\ref{eq:v}) with
$w=0$, $\delta p = 0$ and $\sigma = 0$). In terms of the relative parameterization of Eq.~(\ref{eq:def_eta}) 
we find therefore that
\be
\eta = \frac{2}{3\beta - 1} \label{eq:dgp_eta}.
\ee
Numerically, $\eta$ is small at high redshifts but tends towards $\eta \approx -0.44$ today for a flat
Universe with $\Omega_m = 0.3$. In other words, the gravitational slip $\Pi$ (or equivalently
the effective anisotropic stress) in DGP is of
a size comparable to the gravitational potentials themselves. We should note here that the
extra factors of $1\pm1/(3\beta)$ in Eqs. (\ref{eq:dgp_phi}) and (\ref{eq:dgp_psi}) affect the growth
of the perturbations as well, so that the value of the gravitational potentials is somewhat
different than in a Quintessence model with the same equation of state parameter -- but of course
this is good since it implies that it is possible to measure the extra perturbation level parameters, 
at least if they are of the size found in DGP. 

We have seen that it is very difficult to avoid generating a late-time effective anisotropic stress for the models that
we would call ``modified gravity models'', and in addition the corresponding gravitational slip is generically very large, of the
order of the gravitational potentials themselves. For this reason $\Pi$ is a key diagnostic in the phenomenological
framework: If we find a deviation from $\Lambda$CDM, then constraints on the anisotropic stress can help
to identify a likely explanation: If $\Pi\neq0$ then Quintessence and K-essence type models are ruled
out and a modification of GR looks likely, while in the opposite case modified gravity models are disfavored
and it appears more sensible to look for a minimally coupled field.

The ghost issues in DGP can be cured by adding more higher order terms (taking care to ensure that
the equations of motion stay second order). The most general such theory was worked out in
\cite{springerlink:10.1007/BF01807638}, but only recently have such theories become better
known, like the Galileon \cite{Nicolis:2008in} and similar models like Kinetic Gravity Braiding \cite{Deffayet:2010qz},
massive gravity \cite{deRham:2010ik} and others more \cite{Charmousis:2011bf,DeFelice:2011bh}, described in much more detail elsewhere
in this volume \cite{cras:derham2012}.

Equations (\ref{eq:dgp_phi}) and (\ref{eq:dgp_psi}) have an interesting property: we see that the extra terms cancel 
for the lensing potential $\phi+\psi$. In other words, light is lensed by matter perturbations exactly
as in GR, without any additional lensing. 
This offers another way to see why suppressing the anisotropic stress forces DGP to revert
back to GR: if both $\phi-\psi$ and $\phi+\psi$ are just given by the usual GR expressions without any
dark energy contribution, then the (effective) dark energy does not contain any fluctuations. But
this in general only possible if the dark energy is a cosmological constant: due
to the Bianchi identity and the conservation of the rest of the EMT, also the effective dark energy EMT
is conserved. If $w\neq -1$ the conservation equations (\ref{eq:delta}) and (\ref{eq:v}) couple the
perturbations to the gravitational potentials, and with only one remaining function to choose
($\delta p$) it is in general not possible to find a solution. So not only does $\sigma=0$ suppress
all effective dark energy perturbations for such models, in general it is not even possible to achieve this
self-consistently except for a $\Lambda$CDM-like behavior. Therefore if we want to have a 
non-trivial modified-gravity model with
$\phi=\psi$, then this model needs in general to change the lensing potential.

It is not only DGP that has this property where lensing is unaffected. On the other hand,
the full Galileon case should change lensing \cite{Wyman:2011mp}. I suspect that in this case it may be
possible to construct an explicit example where the anisotropic stress vanishes without
rendering the model unviable -- but this still needs to be demonstrated. But also for
Galileons I expect that in general the anisotropic stress is non-zero. The only way around
the need to fine-tune $\sigma$ to vanish is to include the property already at the level of
the action. I do not know what that condition translates into for a fully general model,
but as mentioned above, for the action (\ref{eq:stact}) the condition is that the coupling to 
$R$ vanishes \cite{Amendola:2012zz}. At least for this class (and for the reason
outlined above) ``anisotropic stress'' and ``modification of gravity'' are synonyms.

On the observational front, current data is consistent with no additional anisotropic stress
(and indeed no detection of any dark energy perturbations), but with errors of order unity
on the additional perturbation parameters $\eta$ (\ref{eq:def_eta}) and $Q$ (\ref{eq:def_Q}) or their equivalents.
Future observations over the next one to two decades will provide much stronger constraints,
reaching about a 10\% accuracy on the perturbation parameters in several redshift and scale bins. For more
details, see for example
\cite{Caldwell:2007cw,Amendola:2007rr,Hu:2007pj,Bertschinger:2008zb,Song:2008vm,Bean:2010zq,Song:2010rm,Pogosian:2010tj,Daniel:2010yt,Song:2010fg,Hojjati:2011ix,Baker:2011jy,Zhao:2011te,Zuntz:2011aq,ECTheory:2012zz}

\section{Limitations and challenges}  \label{sec:problems}

\subsection{The dark degeneracy}  \label{sec:degen}

Any dark components are only detectable through measurements of the left hand side of the Einstein 
equation (\ref{eq:einstein}), and the inferred properties on the energy momentum tensor on the
right hand side. But on the right hand side there is only the {\em total} dark EMT. This means that
a single dark component looks just the same as
\be
T_{\mu\nu}^{\rm (dark)} = T_{\mu\nu}^{\rm (DM)} + T_{\mu\nu}^{(\Lambda)}  + T_{\mu\nu}^{\rm (Q)} 
+ T_{\mu\nu}^{\rm (MG)}  + \ldots
\ee
An immediate consequence is that cosmological data will never be able to tell us that we have
more than one general dark energy component. The only way to reach such a conclusion would
as a probabilistic statement based on model predictions and Occams razor.

But the situation is actually worse: if we restrict ourselves to distance data (or more generally,
data that constrains only the background evolution) and try to reconstruct directly an equation
of state parameter that can fit the data in a model with dark matter and dark energy (a fairly
common approach), we find that
\be
w(z)  = \frac{H(z)^2-\frac{2}{3} H(z) H'(z) (1+z)}{\Omega_m H_0^2 (1+z)^3-H(z)^2} .
\ee
Using a dark energy with this $w(z)$, we will always fit the (background) data which is here given as $H(z)$.
But in this expression $\Omega_m$ is a free parameter that we can choose as we want \cite{Kunz:2007rk}.
We will therefore find a possible dark energy for any amount of dark matter. Indeed, we
cannot even be sure that there is dark matter at all, maybe we are dealing with a single dark
fluid. We can thus generate possible families of dark energy evolutions, parameterized by
$\Omega_m$. Again, only theoretical prejudice or non-gravitational tests can break this
degeneracy. It is for this reason that we plotted only the total $w$ in Fig.~\ref{fig:total_w}.
This is all that we can learn from background data.

The reason is that we tried to measure more degrees of freedom than are present in the
energy momentum tensor. We can only measure one pressure, but instead we try to measure
a general pressure and the amount of dark matter. Of course the degeneracy is broken when
we put constraints on the dark energy pressure, e.g. if we demand a constant $w$. Then 
we can measure both $w$ and $\Omega_m$. But then again, the true dark energy may not
be characterized by a constant $w$, so the extra information that we input into the analysis
may be wrong. The existence of the degeneracy is also a good test for codes
that try to reconstruct a fully general $w(z)$: using only background data there should then
be no constraints on $\Omega_m$.

\begin{figure}[htb]
\begin{center}
\begin{tabular}{cc}
\includegraphics[width=2.55in]{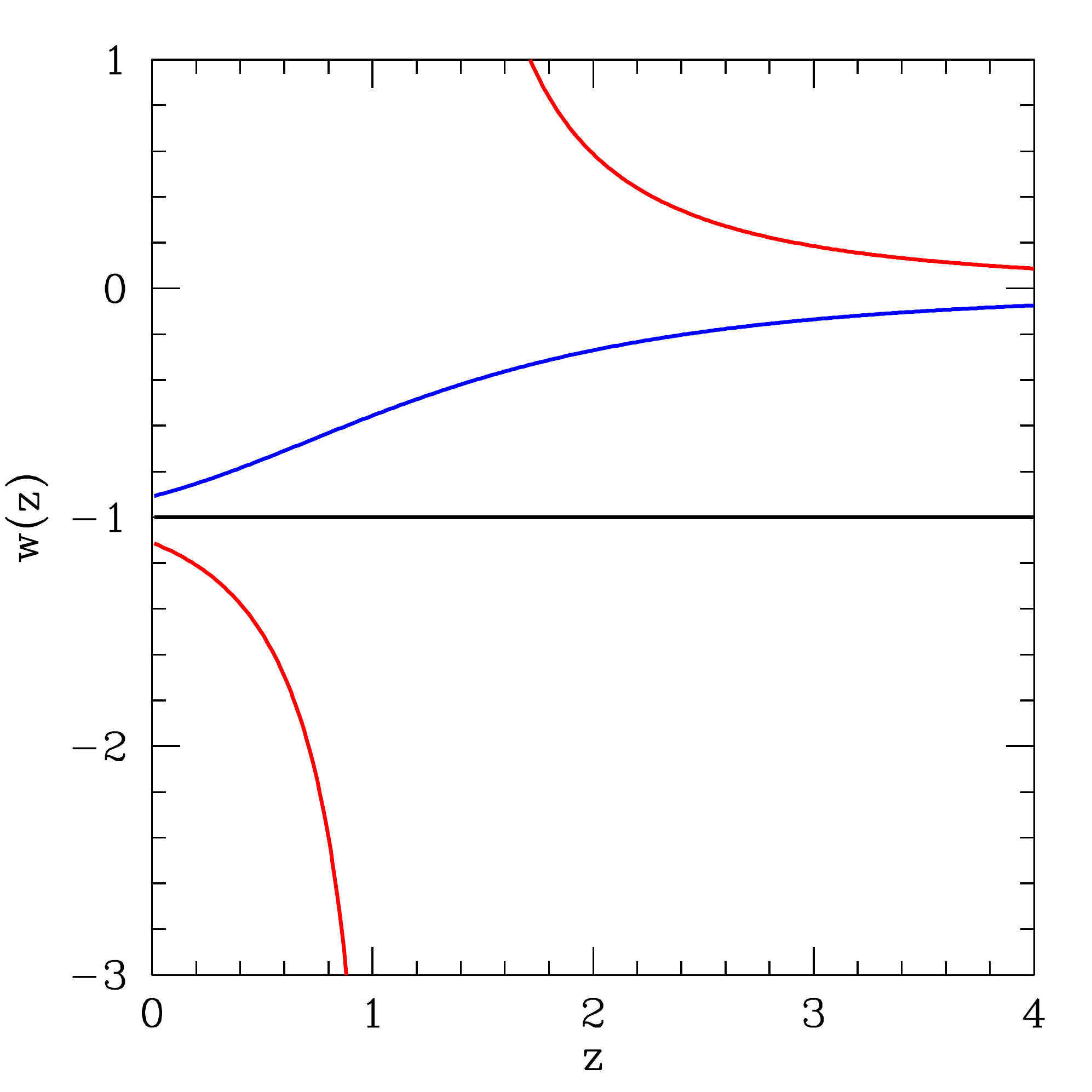}
\includegraphics[width=3.45in]{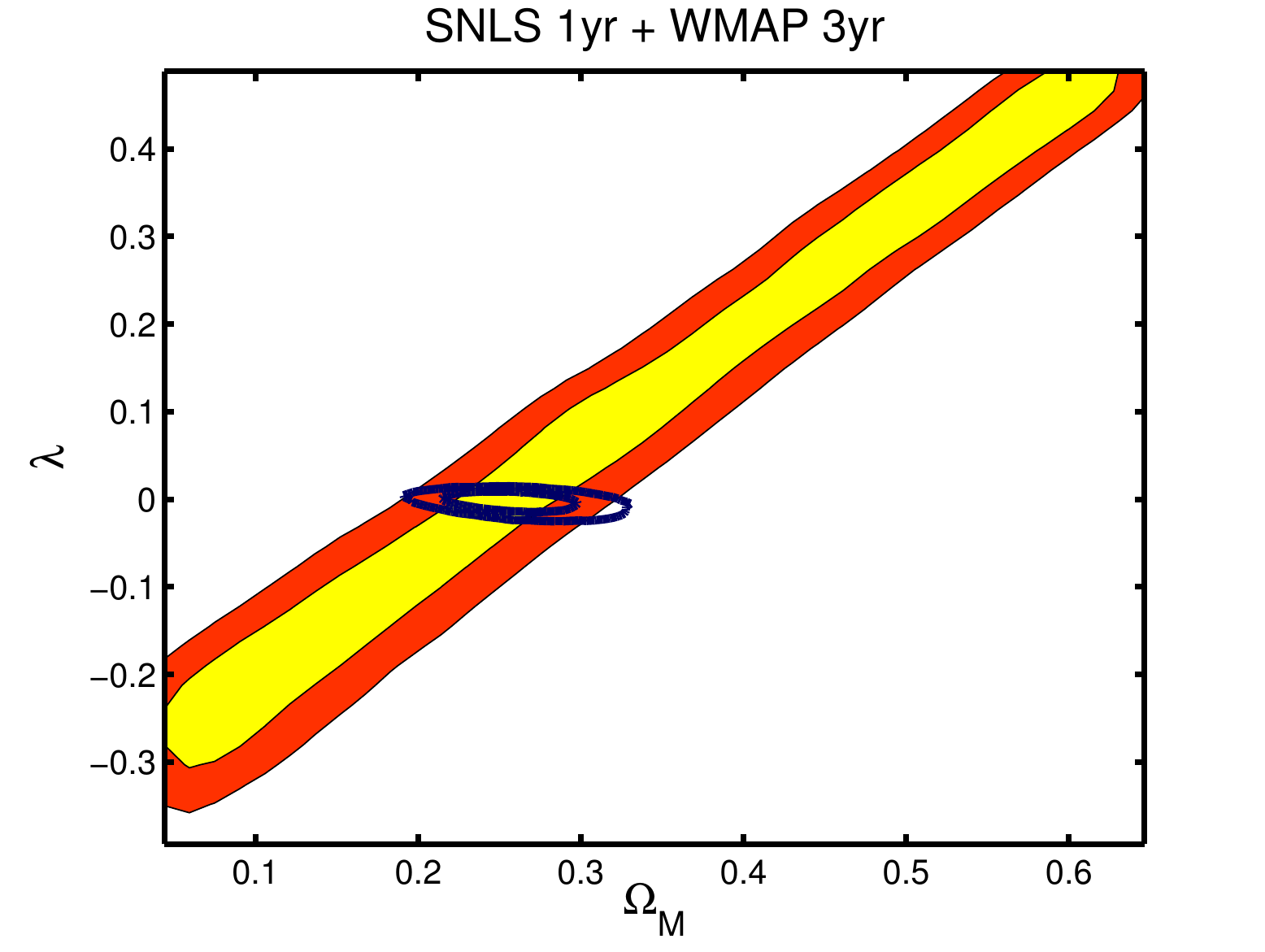}
\end{tabular}
\end{center}
\caption{{\em Left:} Different $w_\lambda(z)=-1/[1-\lambda(1+z)^3]$ with $\lambda=0$ (horizontal black line),
$\lambda > 0$ (red line starting at $w<-1$) and $\lambda<0$ (blue line starting at $w>-1$).
All these $w$'s lead to the same $\Lambda$CDM expansion history if $\Omega_m$ is adjusted accordingly.
{\em Right:} One and two sigma contours for a dark energy model given $w_\lambda(z)$,
$\sigma=0$ and $\cs=0$ (large yellow/red filled contours) and $\cs=1$ (black open contours).
For $\cs=0$ we cannot distinguish between dark matter and dark energy and so cannot
measure $\Omega_m$, while the choice
$\cs=1$ breaks the degeneracy. This example taken from \cite{Kunz:2007rk} uses the first-year 
SNLS supernova data \cite{Astier:2005qq} and the three-year WMAP data \cite{Spergel:2006hy}.
\label{fig:degen}}
\end{figure}

What happens at the level of perturbations? There are two extra functions, $\cs$ and $\sigma$,
that can be constrained. This is the reason why we could constrain in section \ref{sec:quintobs} 
the equation of state $w(z)$ of Quintessence and simultaneously measure $\Omega_m$:
dark matter has $\cs=0$ and Quintessence $\cs=1$. If we tried to do the same with cold
dark energy, characterized by a free $w(z)$ and $\{\cs=0,\sigma=0\}$ then we recreate
the degeneracy between dark energy and dark matter, and in that case it is again impossible
to measure $\Omega_m$, see Fig.\ \ref{fig:degen}. The same is true if we leave $\cs$ and $\sigma$ free: In that case
we are reconstructing the most general dark fluid (within first order perturbation theory and
only considering scalar perturbations), and so any further freedom cannot be constrained
\cite{Hu:1998tj,Kunz:2007nn}.
This is just due to the way Einstein's equations work, and the phenomenological
framework allows at least to diagnose the issue.

We also add more freedom when we introduce couplings between dark matter and dark energy.
Again, since we can only measure the total dark energy momentum tensor, we can always
separate it into an uncoupled sum of a dark matter and a general dark energy EMT,
even if the the true model is actually a coupled model. We can always compute a ``valid''
dark energy EMT through
\be
T_{\mu\nu}^{\rm (DE)} = T_{\mu\nu}^{\rm (total~coupled)} - T_{\mu\nu}^{\rm (DM)} 
\ee
for any choice of dark matter EMT. This means that we can always find an uncoupled dark
energy - dark matter model that gives the same observations as a coupled model. Our only
hope is that the coupled model is sufficiently well motivated on theoretical grounds as well as simpler
to allow accepting it as the correct description. However, we can also do the inverse,
we can always find a coupling function that will allow us to model the dark energy as a cosmological
constant, as long as there is no additional anisotropic stress. To show this explicitly at the background
level, let us assume that we have baryons and a coupled dark matter - dark energy fluid, and
let us write the conservation equations for dark matter and dark energy as
\bea
\dot{\rho}_{\rm DM} + 3 H \rho_{\rm DM} &=& \rho_{\rm DM} C(t) \, , \\
\dot{\rho}_{\rm DE} + 3 H \rho_{\rm DE} &=& - \rho_{\rm DM} C(t)  \, .
\eea
With this, and by {\em assuming} that the dark energy is a cosmological constant and computing $\dot{H}$ 
from the Friedmann equation, we find easily that \cite{Amendola:2012zz}
\be
C(t) = \frac{\ddot{H} - \frac{9}{2} H \Omega_b H_0^2 a^{-3}}{\dot{H} + \frac{3}{2} \Omega_b H_0^2 a^{-3}} + 3 H \, .
\ee
This then is the (averaged) coupling that allows us to recover a cosmological constant
as the dark energy.
Again the problem is that the combination of coupling parameters and a general DE
energy-momentum tensor encompasses more degrees of freedom that can be measured.

Part of the families of $w(a)$ that arise from the dark degeneracy, whether when changing $\Omega_m$
or couplings, contain phantom-crossing and even divergences at low redshift as can be seen in
the left panel of Fig.\ \ref{fig:degen}. This typically happens if we need to
``overcompensate'' the dark energy. However, by construction the expansion rate $H(a)$ and therefore
all observations are unaffected. This illustrates that strange behavior in $w$ and in other
parameters (like the form of the effective sound speed of the Quintom model in figure \ref{fig:quintom}), even divergences,
are not necessarily problematic as long as the metric stays well behaved. Similar effects can apparently
happen for $f(R)$ models \cite{Amendola:2007nt}. Also coupled models can lead to (apparent) phantom
crossing when they are analyzed without taking the coupling into account \cite{Huey:2004qv,Das:2005yj}; 
in this case this is directly the consequence of the dark degeneracy \cite{Kunz:2007rk}.

The dark degeneracy also shows that the dark energy does not need to be smooth at all. Current data
is perfectly compatible with a clustering dark energy with $\cs=0$. It could contain some, or even all, 
of the clustering that is usually attributed to the dark matter. On the other hand, the current data is
indeed compatible also with a smooth dark energy like a cosmological constant or a scalar field model
with high sound speed. For this reason it is tempting to {\em define} the split between dark matter and
dark energy by assigning the clustering part of the dark sector to the dark matter and the rest to the dark 
energy. This works
for simple dark energy models like generalized quintessence since once we have defined the split in
this way, the dark degeneracy guarantees that we can find a $w(z)$ for the dark energy that allows
matching the observed $H(z)$. Although the split would not reflect physical reality if indeed the dark
energy has a low sound speed and does cluster, we are barred by the dark degeneracy to see this
with cosmological measurements alone. But if dark matter experiments one day indicate that the
amount of dark matter is different from the amount observed in a cosmological setting, then we must
not forget that the split of the dark sector into dark matter and dark energy is not unique, and that
a different dark matter density can be accommodated by modifying the dark energy \cite{Kunz:2007nn}.

There is also a cosmological reason why it is a bad idea to use the clustering property directly to define the
dark energy part of the total dark EMT and so break the dark degeneracy: modified gravity models tend to have $Q\neq 1$
in addition to $\eta \neq 0$. The non-zero anisotropic stress in these models contributes to the clustering through the
last term in Eq.~(\ref{eq:v}). By defining the dark energy to be smooth, we
would exclude some of the models that we want to test with help of the data.

A closely related degeneracy exists between curvature and the dark energy
\cite{Hlozek:2008mt,Clarkson:2007bc}. Since the global curvature contributes only to the
background expansion rate, it is easily confused with a smooth dark energy component.
In order to break the curvature degeneracy, one needs to use a geometrical test for the
presence of non-zero curvature, for example by comparing longitudinal and transverse
distances \cite{Clarkson:2007pz}. So at least in principle it is possible to break the
curvature degeneracy with cosmological observations (while the dark degeneracy
is in principle perfect and unbreakable) but until this has been done one needs to be
very careful not to mistake a small curvature contribution as a smooth, evolving dark
energy.

\subsection{Non-linear evolution}  \label{sec:nonlinear}

While the fundamental principle of measuring the geometry and then using Einstein's equations
to map the results to an effective EMT (or other effective parameters) holds at the fully non-linear
level, we have been working mostly with linear perturbation theory\footnote{It is worth noting that the split of the energy-momentum tensor into an energy density, an
isotropic pressure, an energy flux and an anisotropic stress can be defined at the fully non-linear
level with the help of the covariant 1+3 formalism, see e.g.\ \cite{Maartens:1998xg}.}. However, in the future we will
want to consider at least mildly non-linear scales to make the best use of the data.

At the non-linear level we can expect some surprises to show up. Here just one example:
when we look at the sky, we find that galaxies (and thus likely also the dark matter) are not static,
but they move around. Such a motion induces a thermal pressure and anisotropic stress at second
order {\em even for pressureless dark matter} \cite{Ballesteros:2011cm},
\be
p_m = \frac{1}{3} v_m^2 \rho_m \, , \quad  \pi_m^{\mu\nu} = \rho_m v_m^{\langle\mu}v_m^{\nu\rangle} \, .
\ee
(The combination $\pi_m^{\mu\nu}$ is the projected symmetric trace-free part,
the fluid anisotropic stress $\sigma$ is related to its scalar part; see the appendix B.1 of \cite{Ballesteros:2011cm} 
for the precise definitions.) This contribution is present even in $\Lambda$CDM.

Since we can only observe the total dark EMT, we cannot distinguish the higher-order contributions
to the dark matter pressure and anisotropic stress from the fundamental properties of the dark energy,
another manifestation of the dark degeneracy of section \ref{sec:degen}. We can of course also not
choose to measure only the linear perturbations, we always observe the fully non-linear result. We
have to be very careful not to mix up such a non-linear contribution with the actual dark energy properties.
Otherwise there is a risk that we will wrongly claim a detection of dark energy perturbations, ruling
out $\Lambda$CDM by mistake. However, at least on linear scales the peculiar velocity contribution
to $\delta p$ and $\sigma$ is
tiny and probably negligible \cite{Ballesteros:2011cm}.

\subsection{Dependence on the environment}  \label{sec:environment}

A further issue relevant for the phenomenological description concerns a possible
dependence on local physics. For example, a screening mechanism like the Vainshtein \cite{Vainshtein:1972sx}
or Chameleon \cite{Khoury:2003aq,Brax:2004qh,Hu:2007nk} mechanism, based on local quantities like
the density is an integral part of all viable MG models. Without such a screening these models all
fall foul of local tests of gravity that place very strict limits on fifth forces in the solar system. But
such a mechanism is difficult to model with the phenomenological description as it is not purely
a function of scale $k$ \cite{Zhao:2011cu}, and a description at the level of the action that
incorporates the screening effect may be preferable \cite{Brax:2011aw}. 
On the other hand, power spectra are also written as a function of scale and so phase the
same problem. Environmental dependence is
therefore a general concern for data analysis as a whole.

The precise way in which a given physical modified gravity model makes the transition from
to the small-scale regime where the modifications are suppressed is actually an important
discriminant: The action given in (\ref{eq:stact}) already contains the possible degrees of
freedom to first order in perturbation theory. Measuring the evolution of the Universe on
large scales will allow to constrain the functions $f(\varphi)$ (through the anisotropic
stress), $V(\varphi)$ (mostly through the background expansion) and $K(X)$ (mostly through
the sound speed). But this is certainly not the only action with these degrees of freedom that can be written down, and 
there is a degeneracy between different actions that can at least partially be broken by
considering the additional information from the way the model reverts to GR on small scales.

\section{Conclusions and Outlook}  \label{sec:conclusions}

In this review we have discussed a general framework that allows to parameterize all possible
outcomes of cosmological observations.\footnote{Small print: At least to first order and for scalar perturbations.
The latter is straightforward to rectify by including vector and tensor perturbations, while extensions
to non-linear perturbations present a more formidable challenge, at least concerning the interpretation
of the results.}
We found that this can be done in different ways that
are all directly related but may be better in some situation or other. The first approach is to
use an effective dark energy momentum tensor, where the relevant degrees of freedom can be
taken to be the average (background) pressure $p(t)$, and the pressure perturbation $\delta p(t,k)$ as
well as the anisotropic stress $\sigma(t,k)$. Usually the background pressure is given in terms of
an equation of state parameter $w=p/\rho$, and often the pressure perturbation is parameterized 
by a rest-frame sound speed $\cs$.

The Einstein equations relate these quantities to the geometry, where the equivalent degrees of freedom
are the average expansion rate $H$ and (in the conformal Newtonian gauge) the gravitational potentials
$\phi$ and $\psi$. There are many ways to introduce other parameterizations, a common approach is
to use $w$ and in addition parameters that quantify the deviation from a hypothetical universe where
the gravitational potentials are given by the perturbations in the matter (and radiation) alone. The main 
point here is that one always needs {\em one} background parameter and {\em two} perturbation level parameters
for all measurements that involve perturbations, like the CMB, galaxy clustering, weak lensing, redshift
space distortions, etc. Conversely, these 1+2 functions also represent the information that can be
extracted from measurements.

In this way we have built a framework that allows to combine all possible cosmological observations,
by parameterizing exactly the possible degrees of freedom. With this, we are also sure that we can
probe the consistency of observations with the still favored $\Lambda$CDM model in all possible ways.

We have further seen that the phenomenological parameters can be mapped to functions in an action
that describes gravity and the fields present in the Universe. Through this we can reconstruct actions
from the observations, but again the limited information available will mean that there are degeneracies
that cannot be broken. Nonetheless the phenomenological parameters will give an indication
of which classes of models we should investigate in more detail. As an example, the presence of a strong
late-time anisotropic stress appears closely related to a modification of gravity. 

Nonetheless, the framework rests on the validity of the cosmological principle, i.e. statistical isotropy
and homogeneity. If the Universe has a fundamentally different structure, e.g. if we lived in the center
of a gigantic void, then further degrees of freedom become possible. As an example, in such a case
there are two different functions that parameterize the background evolution, that can be taken to be
a transverse and a longitudinal expansion rate. In a FLRW background they are equal, but in a
Lema\^\i tre-Tolman-Bondi (LTB) background they are in general different. It has been shown that in this
case they can be distinguished by measurements of $H(z)$ and distances \cite{Clarkson:2007pz,Uzan:2008qp}.
The full perturbation theory for the LTB background is still being worked out, and there are of course
further generalizations possible. Hence when using the phenomenological approach one should always
be aware of the FLRW assumption that has been made, and one should test this assumption as far
as possible.

Today the phenomenological parameters are not well constrained, with deviations of order unity
still possible. However, future very large surveys like {\sc Euclid} \cite{EditorialTeam:2011mu}
will significantly tighten the bounds and will permit to constrain $w$ at the 1\% level
and the perturbative quantities at the 10\% level \cite{Caldwell:2007cw,Amendola:2007rr,Hu:2007pj,Bertschinger:2008zb,Song:2008vm,Bean:2010zq,Song:2010rm,Pogosian:2010tj,Daniel:2010yt,Song:2010fg,Hojjati:2011ix,Baker:2011jy,Zhao:2011te,Zuntz:2011aq,ECTheory:2012zz}. To reach this precision will
still take a lot of work to control the systematic errors in the observations, but also to ensure that all
important theoretical effects are taken into account, for example non-linear behavior of the perturbations, 
but also relativistic effects \cite{Yoo:2010ni,Bonvin:2011bg,Challinor:2011bk,Jeong:2011as} as well
as including vector and tensor modes.

\acknowledgments

I would like to thank Philippe Brax and C\'eline Boehm 
for inviting me to write this mini-review. 
It is a pleasure to thank  all my dark energy collaborators over the years, 
and I am especially grateful to Bruce Bassett, Philippe Brax, Ruth Durrer, Lukas Hollenstein, Dragan Huterer 
and Ignacy Sawicki for helpful discussions and comments about this manuscript.
This work is supported by the Swiss National Science Foundation.
I gratefully acknowledge hospitality by UNFPA Thailand during part of the writing of this review.

\bibliography{cras_mk}
\bibliographystyle{JHEP}

\end{document}